\documentclass[3p]{elsarticle} 
\makeatletter
\def\ps@pprintTitle{%
 \let\@oddhead\@empty
 \let\@evenhead\@empty
 \def\@oddfoot{\centerline{\thepage}}%
 \let\@evenfoot\@oddfoot}
\makeatother

\date{}
\def\texpsfig#1#2#3{\vbox{\kern #3\hbox{\includegraphics{#1}\kern #2}}\typeout{(#1)}}

\usepackage{url,hyperref}
\usepackage{bbm}
\usepackage{eurosym}                           
\usepackage{url}                               
\usepackage{longtable}                         
\usepackage{array}                             
\usepackage{graphicx,color}
\usepackage{amsthm}
\usepackage{amsbsy}

\usepackage{amssymb}
\usepackage{amsmath}
\usepackage{enumerate}
\usepackage{graphicx,color}
\usepackage{epsfig}
\usepackage[ruled,vlined]{algorithm2e}
\usepackage{multirow,bigdelim}
\usepackage[table]{xcolor}
\usepackage{natbib}
\theoremstyle{plain}
\newtheorem{thm}{Theorem}[section]

\newtheorem*{rem}{Remark}
\theoremstyle{remark}

\theoremstyle{plain}
\newtheorem{lem}[thm]{Lemma}

\theoremstyle{definition}

\newcommand{\e}{{\rm e}}        
\def\R{\mathbb{ R}}             
\def\E{\mathbb{ E}}             

\def\P{\mathbb{ P}}             

\def\F{\mathcal{F}}             
\def\corr{\text{corr}}            

\def\Var{\mathbb{V}\text{ar}}   
\renewcommand{\d}{{\rm d}}      
\def\dW{{\rm d}W}               
\def\dt{{\rm d}t}

\def\dx{{\rm d}x}

\def\1{{\mathbbm{1}}}            

\theoremstyle{plain}

\usepackage[margin=1cm]{caption}

\numberwithin{equation}{section}	     

\newcommand{\zbox}[1]{
\noindent
\begin{center}
\framebox[14.5cm]{
\begin{minipage}{14cm}
#1
\end{minipage}
}
\end{center}
 }

\title{Efficient Pricing and Calibration of High-Dimensional Basket Options}

\begin{document}

\author[1,2]{Lech A.~Grzelak\corref{cor1}}
\ead{L.A.Grzelak@uu.nl}
\author[3,4]{Juliusz Jablecki}
\ead{J.Jablecki@uw.edu.pl}
\author[5]{Dariusz Gatarek}
\ead{Dariusz.Gatarek@ibspan.waw.pl}
\cortext[cor1]{Corresponding author.}

\address[1]{Financial Engineering, Rabobank, Utrecht, the Netherlands}
\address[2]{Mathematical Institute, Utrecht University, Utrecht, the Netherlands}
\address[3]{Financial Risk Management Department, National Bank of Poland, Warsaw, Poland}
\address[4]{Chair of Quantitative Finance, Faculty of Economic Sciences, University of Warsaw, Warsaw, Poland}
\address[5]{Systems Research Institute, Polish Academy of Sciences, Warsaw, Poland}

\begin{abstract}
    \noindent \noindent

    This paper studies equity basket options -- i.e. multi-dimensional derivatives whose payoffs depend on the value of a weighted sum of the underlying stocks -- and develops a new and innovative approach to ensure consistency between options on individual stocks and on the index comprising them. Specifically, we show how to resolve a well-known problem that when individual constituent distributions of an equity index are inferred from the single-stock option markets and combined in a multi-dimensional local/stochastic volatility model, the resulting basket option prices will not generate a skew matching that of the options on the equity index corresponding to the basket. To address this ``insufficient skewness'', we proceed in two steps. First, we propose an ``effective'' local volatility model by mapping the general multi-dimensional basket onto a collection of marginal distributions.    Second, we build a multivariate dependence structure between all the marginal distributions assuming a jump-diffusion model for the effective projection parameters, and show how to calibrate the basket to the index smile. Numerical tests and calibration exercises demonstrate an excellent fit for a basket of as many as 30 stocks with fast calculation time.

\end{abstract}

\maketitle

\section{Introduction}
Basket options are contingent claims in which the underlying is a group of assets -- typically equities (single stocks as well as equity indices/ETFs), commodities or currencies. As with standard options, the holder of a basket option has the right, but no obligation, to buy (call) or sell (put) the group at a specified strike price which itself is based on the weighted value of the component assets. Basket options can be traded on their own, but commonly feature implicitly in structured products like equity basket linked notes (ELNs) or equity basket certificates of deposit, where return to investors depends on the percentage change in the value of stocks or stock indices in the basket, with partial or total capital protection. For example, on April 16, 2021, JP Morgan priced $\$$9.6 million of notes linked to a basket of 30 unequally weighted U.S. stocks with exposure to infrastructure\footnote{Cf. the prospectus availible in the online records of the U.S. Securities and Exchange Commission at: https://www.sec.gov/Archives/edgar/data/19617/000089109221003578/e13291-424b2.htm}. The structure is effectively a combination of a European-style basket option and a hypothetical zero coupon bond of the issuing entity.

The key feature of basket options, and any structured notes based on them, is their inherently multidimensional risk profile which presents two sets of related challenges. First, there is the obvious computational problem of handling a potentially large number of correlated risk factors driving the basket, which in turn requires involved Monte Carlo simulation and calls for the deployment of some dimension-reduction technique, especially in light of the fact that many such structures exhibit early exercise or other exotic features (e.g. Asian or lookback options).

Second, and perhaps more importantly, there is the problem of ensuring consistency between model dynamics chosen for the basket and its constituents, on the one hand, and market pricing on the other. Typically, stocks or stock indices/ETFs included in the basket will have their own vanilla options markets and the basket collection itself can also be traded on its own account. Thus, ensuring consistency entails not only fitting the implied volatility smiles/skews for individual assets in the basket -- which can be done with excellent precision using e.g. the local volatility model or the Heston model -- but also their covariance structure, such as to match closely the implied volatility smile for the market index corresponding to the basket.

Market practice is to use the local volatility model for individual basket members and historical correlations estimates between underliers to relate the respective risk factors in a Monte Carlo simulation.\footnote{For example, in the Bloomberg basket options pricing template correlations are, by default, estimated over a 5 year period, whereby to eliminate noise, a given percentile of rolling 6-month cross-correlation estimates is chosen in the parameterization of the full correlation matrix.} In addition, to avoid situations in which the correlation matrix fails to be positive semi-definite, a flat correlation level is often imposed. However, this procedure does not automatically ensure that the model price of a basket option is consistent with the market price of the option on the index corresponding to the basket. And even if, as per market practice, the correlation coefficient is chosen specifically to match the price of at-the-money index options, the procedure generates an implied volatility smile/skew that for the basket that is significantly less pronounced than that observed for the index options.

\begin{figure}[!h]
    \centering
    \includegraphics[width=1.0\textwidth]{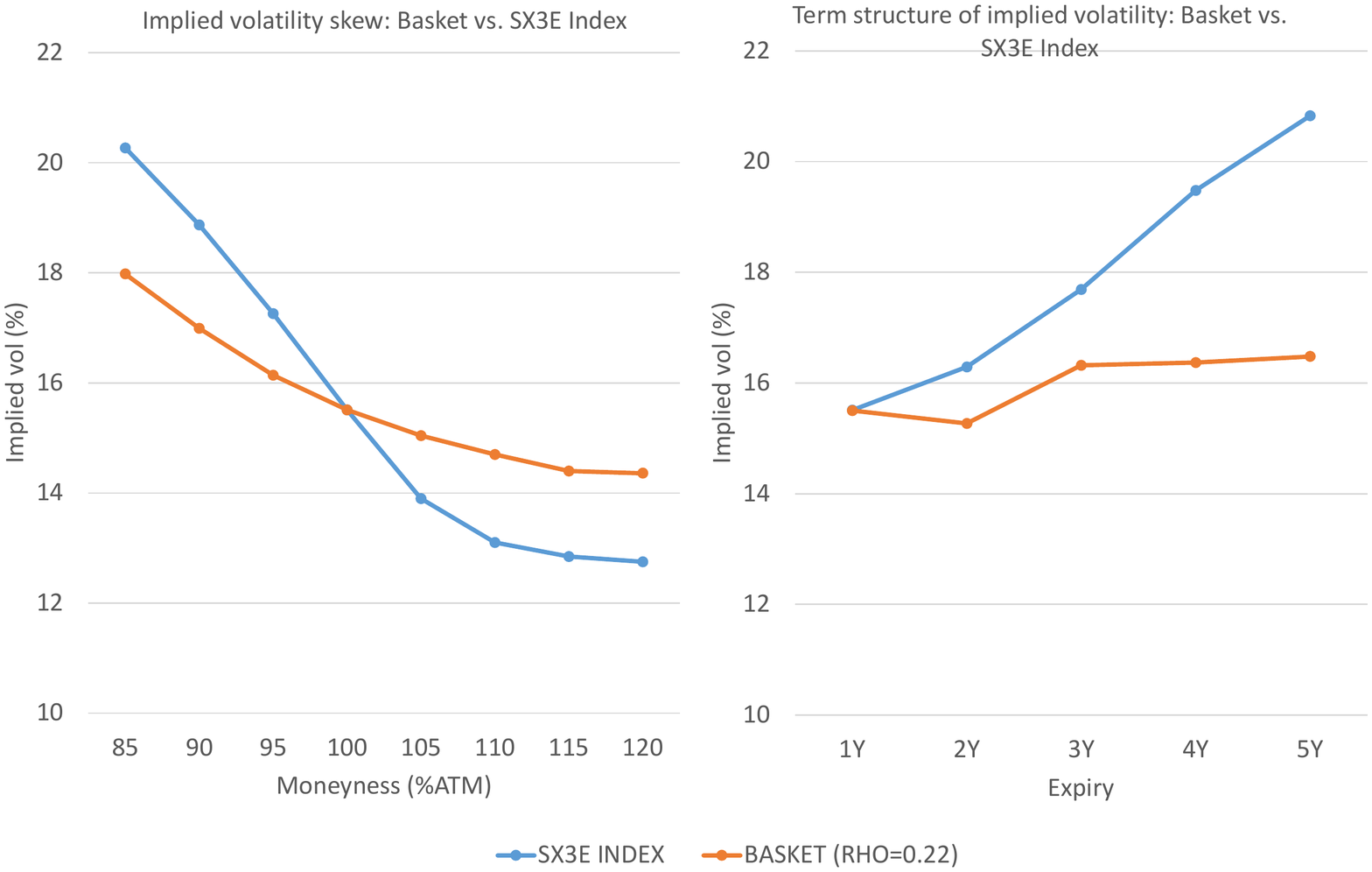}
    \caption{Local volatility model calibrated to Euro STOXX Food and Beverages Index (SX3E) and a corresponding basket of its constituents; LHS: skew for 1Y index options; RHS: implied volatility term structure of index options; data as of 1 April 2022; correlation parameter $\rho=0.22$ calibrated to 1Y ATM implied SX3E volatility.}
    \label{fig:mkt_smile}
\end{figure}

Consider for example options on the EURO STOXX Food and Beverage Index (Bloomberg ticker SX3E). The index currently comprises 11 large cap European equities from the food and beverage sector, such as French Danone, Dutch Heineken and Irish Kerry Group, each of which has its own listed options market. Matching 1Y ATM implied volatility for the index (15.51$\%$) requires setting $\rho=0.22$ (data as of 1 April 2022). Repricing the basket with this flat correlation parameter for different moneyness levels generates an implied volatility skew of about 3.6 vol points,\footnote{We define the skew here loosely as the difference in implied volatilities between the 85-120$\%$ ATM levels.} which is less than half of the skew observed in the market (7.5 vol points; cf. Figure~\ref{fig:mkt_smile}, LHS). Similarly, fitting correlation to 1Y ATM vol does not allow to reproduce the term structure of index implied volatilities observed in the market (Figure \ref{fig:mkt_smile}, RHS).

The above heuristic argument shows that a covariance structure implied by the modeling choice, like the multi-dimensional local volatility (or perhaps Heston) model, will not in general match the market well. This pattern becomes even more pronounced when handling really large baskets of tens or hundreds of dimensions, as the generated implied volatilities degenerates to almost a flat line, resembling implied volatilities produced by a simple Black-Scholes model. These shortcomings have been known in the literature for quite some time, as a number of papers have documented that when a constant correlation is picked to match the price of the ATM implied volatility of the index, it generates a skew that is smaller by roughly a factor of two than the market skew -- as indeed shown in Figure \ref{fig:mkt_smile} (see \cite{Bakshi:2003}, \cite{Branger:2004}, and \cite{Bollen:2004}).

One could argue at this point that since the implied volatilities for the index are generally available, there is no need to model individual underlying assets, and the problem of pricing a basket corresponding exactly to the index is a theoretical exercise with little practical relevance. Note, however, that by building a model for the basket that matches the covariance structure implied by index options (while simultaneously calibrating to individual stocks' options) we can easily price products whose value depends on \textit{any} linear combination of \textit{any} subset of stocks included in the broader index. Hence, in essence, we propose a tool for transferring market-implied covariance patterns between liquid and illiquid baskets. This allows us, for example, to price a basket of utilities stocks included in FTSE 100 or health care stocks in Dow Jones, while making sure that the prices of such illiquid baskets will be consistent with the covariance structures implied by prices of the respective liquid index options.

Several approaches have recently been proposed in the literature to tackle the problem of pricing basket options and their skews. Perhaps most notably, so called local correlation models -- which come in various shapes and forms -- allow correlation between stocks to depend on their prices, just like in standard local volatility models (\cite{langnau2009introduction}, \cite{langnau2010dynamic}, \cite{Reghai2010}). Local correlation offered the first consistent remedy to the problem of insufficient skewness referred to above (cf. Figure \ref{fig:mkt_smile}; \cite{Bakshi:2003}), however not without a cost. Local correlation models tend to be slow, difficult to implement and lead to the usual problems of ensuring a positive semi-definite correlation matrix, all of which limits their practical usefulness in a truly multidimensional setting.

A related family of models proposes stochastic instantaneous correlation, usually introduced via Jacobi processes either in a pure form or with jumps (see e.g. \cite{AhdidaA2013} and \cite{Zetocha2015}). Such models aim not only to fit the market prices for correlation products but also, more specifically, to address the so called correlation skew, i.e. an empirically documented tendency for cross-correlations to rise when the market falls. Again, however, the added complexity leads to a further increase in dimensionality and the associated computational burden.

Still others propose to enforce the desired level of the skew through non-Gaussian copulas (e.g. \cite{Lucic2013}), however the resulting model is ``black boxy'' in nature and lacks dynamics, which means it can only be applied to products with a single maturity, as observed in \cite{Zetocha2015}. Moreover, finding a multi-dimensional copula that isn't Gaussian can be challenging in practice, and anyway the need to calibrate the model through brute-force Monte Carlo makes the whole approach slow and costly.

Finally, \cite{Fonseca2006} proposes an analytical framework to handle stochastic correlation via Wishart processes while \cite{langnau2011marking} generalizes Merton’s jump diffusion option pricing approach to the multi-asset case, however without offering a comprehensive solution that would match both the basket skew as well as marginals of the constituent stocks.

Against such a background, we try to contribute to the existing literature by building a highly efficient numerically and robust model capable of handling even very large baskets while allowing quick and precise calibration. In terms of structure, the model is a local volatility type and arises by mapping the basket on a set of proxy variables ensuring that both marginal distributions for individual stocks are matched \textit{and} the covariance structure of the basket corresponds to that of the traded index. Specifically, to ensure excellent fitting of individual stocks' marginal distributions, we leverage the so-called Stochastic Collocation Monte Carlo Sampler (cf. \cite{Grzelak:2015SCMC}), which is a computationally cheap method for approximating expensive distributions (we make no assumptions as to the models driving the respective stocks in the basket). We then control the overall covariance structure of thus reconstituted basket via a judicious choice of ``kernel'' processes, whereby -- similarly as in \cite{langnau2011marking} -- we utilize Merton's jump-diffusion model \cite{merton1976option} and show that the mean and volatility of Poisson jumps, together with cross-correlations of individual kernel processes, produce a rich enough dependence structure to fit both ATM index vols and skew with excellent precision. The proposed approach can be extended into a fully stochastic volatility setting; however, as we shall see, there appears to be no need for the added complexity as the classic Merton diffusion already delivers acceptable results.

The rest of the paper proceeds as follows. Section \ref{modeling_overview} presents the general idea of the model and discusses the stochastic collocation approach to basket reconstitution. Section \ref{sect_calibration} discusses calibration strategy followed by a discussion of numerical exercises in Section \ref{sect_numerical}. Finally, Section \ref{conclusions} draws conclusions.

\section{The Modeling Framework}\label{modeling_overview}
We consider a collection of $N$ assets\footnote{Without loss of generality, we shall henceforth think of the underlying assets as stocks, however the method developed below is obviously general and, mutatis mutandis, applies to other instruments as well.} $S_1(t),S_2(t),\dots,S_N(t)$ and define a basket, $B(\cdot)$, as consisting of $w_i$ shares of each individual stock $S_n(t)$, such that at any time $t$ its price is given by:
\begin{eqnarray}
\label{eqn:basket}
B(t)=\sum_{i=1}^N\omega_iS_i(t),\;\;\;\omega_j\in\R.
\end{eqnarray}
In what follows, for modeling purposes, we shall distinguish between a basket $B(\cdot)$ as a portfolio of stocks, and a stock index, with price process $I(t)$, representing the underlying instrument for equity index options (e.g. an ETF), with realistic values of $N$ ranging from 30-50 (DAX Index) up to hundreds (S\&P 500 Index) or even thousands constituents (Russell 3000 Index).

Our approach to valuing structured payoffs on a basket $B(t)$ will be endowed with three critical features: (i) low dimensionality to facilitate quick and efficient pricing; (ii) consistency with the values of options on individual stocks making up the basket; and (iii) consistency with market prices of options on the stock index corresponding to the basket.

We pursue features (i)-(iii) above by building a one-dimensional local volatility process~\footnote{The proposed framework can also be extended with a stochastic volatility process. Such an extension is trivial and will, for simplicity, be omitted.}:
\begin{eqnarray}
\label{eqn:LV_S}
\d \bar B(t) = r \bar B(t)\dt + \sigma_{LV}(t,\bar B(t))\bar B(t)\dW(t),
\end{eqnarray}
with $r$ being an interest rate and $\sigma_{LV}$ the local volatility function, defined as follows:
    \begin{eqnarray}
    \small
\label{eqn:LV_sigma}
	\sigma_{LV}^2(t,k)=\frac{\frac{\partial }{\partial t}\left[\e^{-rt}\int_k^\infty(y-k)f_{B(t)}(y)\d y\right]+rk\left(F_{B(t)}(k)-1\right)}{\frac{1}{2}k^2_if_{B(t)}(k)},
        \normalfont
\end{eqnarray}
where $F_{B(t)}(\cdot)$ is the CDF of the basket $B(t)$ and $f_{B(t)}(\cdot)$ the corresponding PDF. We shall require that the approximate model for the basket $\bar B(t)$ generates an implied volatility smile/skew matching the implied volatility surface observed in the market for options on the corresponding stock index. This will ensure that prices of index options derived from our basket model will by definition match perfectly (cf. \cite{Dupire:1994},~\cite{Dupire:1996:AUnifiedTheory},~\cite{gyongy86}). Furthermore, it will also guarantee consistency between model-derived prices of baskets on any subset $\{S_{i_1},...,S_{i_K}\} \subseteq \{S_1,...,S_N\}$ of the $N$ stocks included in the index and the prices of liquid index options.

In practical terms, our approach entails two critical steps. First, for each $T_j$ we reconstitute the basket $B(T_j)$ by projecting each process $S_i$ -- assumed to be known only through its marginal distributions/options prices -- on a polynomial of synthetic variables $g_{i,j}(X_i(T_j))$, such that
\begin{eqnarray}
\label{eqn:BasketProxy}
S_i(T_j)\approx\e^{g_{i,j}(X_i(T_j))},\;\;\;\text{and}\;\;\;B(T_j)\approx \sum_{i=1}^N\e^{g_{i,j}(X_i(T_j))}.
\end{eqnarray}
The construction of $g_{i,j}(\cdot)$ will be described in detail below, but its main role will be to ensure the fit of marginal distributions for any choice of $X_i(T_j)$. The reconstitution of the basket paves the way for the second stage of the modeling procedure, namely imposing a multivariate dependence structure on the surrogate kernel processes such as to match the index implied volatility skew, or equivalently, matching marginal distributions between the basket and the index for each maturity point $T_j$, i.e. ensuring that: $B(T_j)\stackrel{d}{=}I(T_j).$ This step will be handled through parameters of the ``kernel'' processes $X_j(t)$, which we assume to be driven by Merton-type jump diffusions. This procedure will allow us to both calibrate the model to individual stocks, and control/modify the covariance structure of all the underlying assets.

\subsection{Covariance Structure of the Basket and the ``Leaking Correlation'' Problem}\label{sec:calibration}
At this stage it may be worthwhile to pause for a moment and provide some further motivation for why controlling basket dependence structure is so tricky and a judicious choice of kernel processes is required. After all, it could be argued that the model price of a basket option
can be fitted to a desired market level by altering the correlations $\rho_{i,j}$ between the Brownian motions driving individual stocks $S_i,S_j$. Such correlation coefficients would then essentially become model inputs describing the stochastic nature of the underlying assets. However, as discussed in~\cite{Andersen:2008:EfficientSimulation}, correlations can be expected to ``leak'', i.e. dissolve, in a Monte Carlo setting, so that effective correlations between the respective basket constituents ends up weaker than assumed. Clearly, this will limit the extent to which we may be able to control the basket smile generated by our model, leading to erroneous pricing of derivatives away from at-the-money levels.

\begin{figure}[htb!]
\centering
 \includegraphics[width=0.48\textwidth]{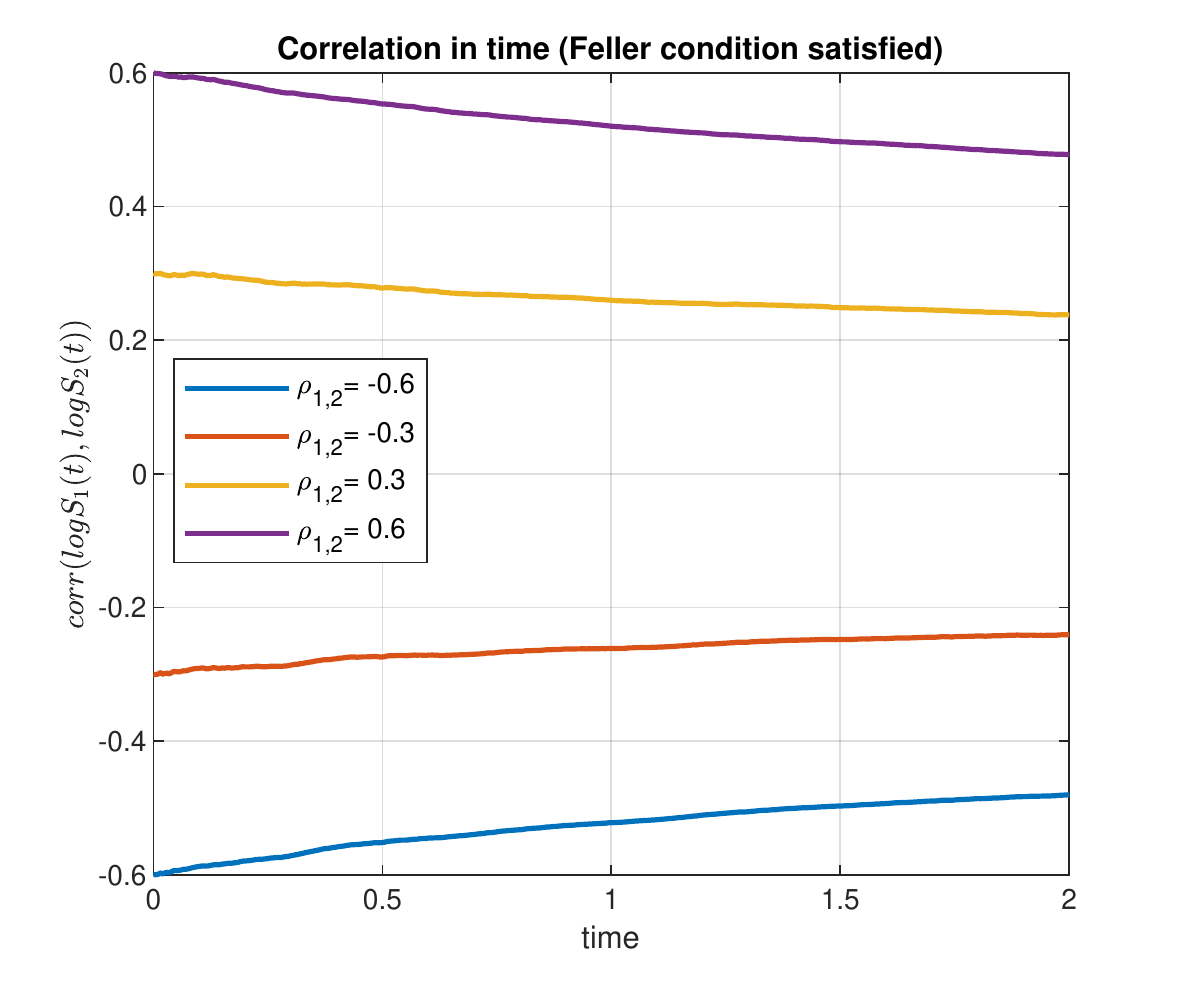}
 \includegraphics[width=0.48\textwidth]{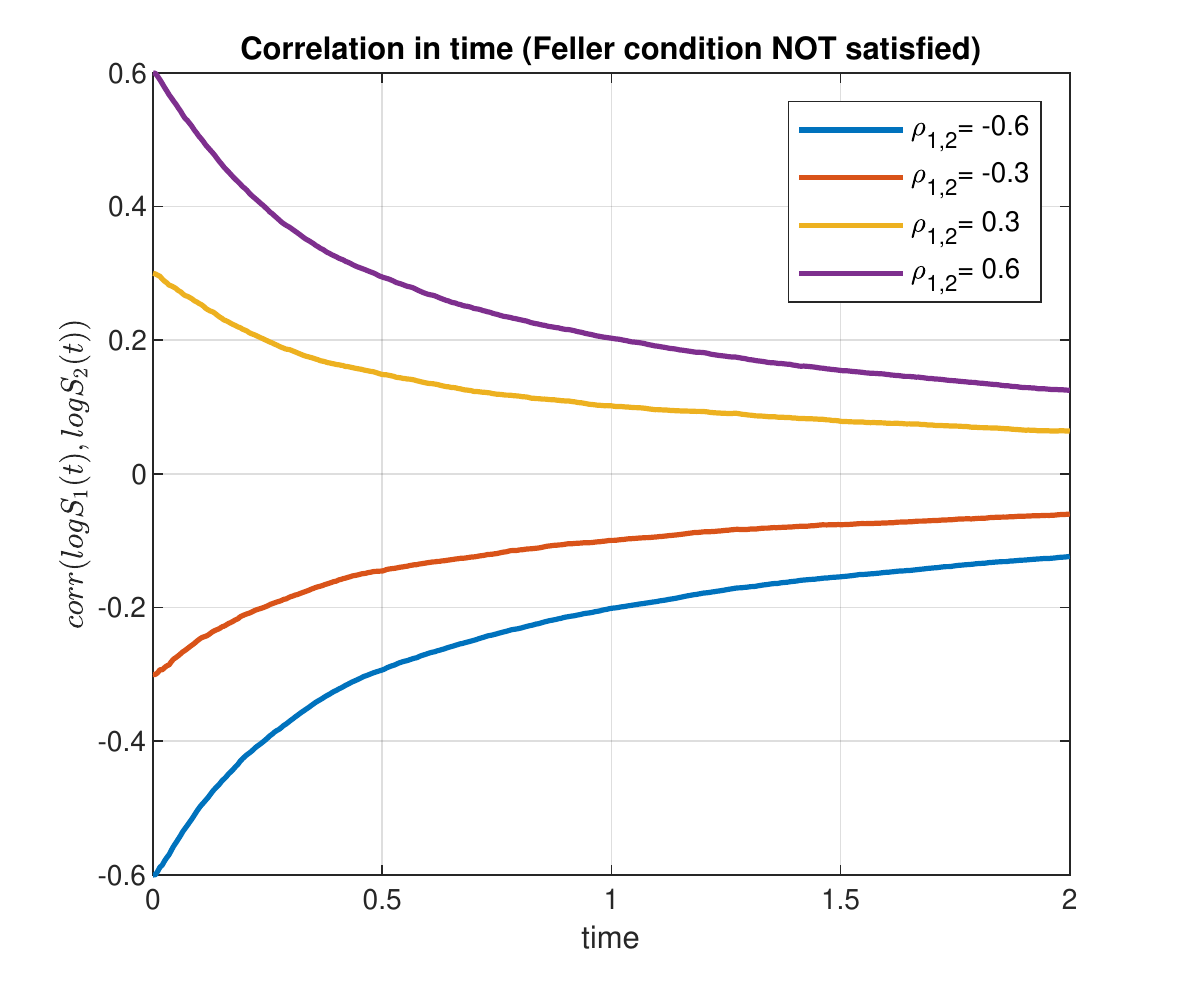}
\caption{Correlation, $\corr(\log S_1(t),\log S_2(t))$, as a function of time. Left: Feller condition satisfied; Right: Feller condition not satisfied. }
\label{fig:leaking_correlation}
\end{figure}

To illustrate the correlation leakage problem, consider two stock exchange processes $S_1(t)$ and $S_2(t)$ driven by the Heston model with correlated Brownian motions\footnote{The respective dynamics are given by ($j=1,2$): $\d S_j(t)=rS_j(t)\dt+v_j^{1/2}(t)S_j(t)\dW_{j,1}(t)$,
$\d v_j(t)=\kappa_j(\bar{v}_j- v_j(t))\dt+\gamma_jv_j^{1/2}(t)\dW_{j,2}(t)$ with correlations $\dW_{j,1}(t)\dW_{j,2}(t)=\rho_{j}\dt$, $\dW_{1,1}(t)\dW_{2,1}(t)=\rho_{1,2}\dt$ and $\dW_{j,2}(t)\dW_{k,2}(t)=0\cdot\dt$. For reference, we set  $S_1(t_0)=1,$ $S_2(t_0)=2.5,$  $r=0,$ $\kappa_1=1,$ $\kappa_2=0.5,$ $\gamma_1=1,$ $\gamma_2=0.6,$ $\rho_{S_1,v_1}=-0.5$, $\rho_{S_2,v_2}=-0.7$, $v_{1,0}=0.1$, $v_{2,0}=0.05$, $\bar{v}_1=0.1$ and $\bar{v}_2=0.05.$}. In the simulation of the underlying processes, we consider a simple Euler discretization as in~\cite{OosterleeGrzelakBook}. In Figure~\ref{fig:leaking_correlation} the cases for the Feller's condition are presented for a range of correlation values. We see that the imposed correlations are not preserved in time, even when Feller's condition is satisfied.
As a potential remedy to ``leaking correlation'', we could attempt to enforce the desired dependence patterns via copula. This, however, does not guarantee success either. To see why, consider the previous example of the Heston model but now -- to make the exposition clearer -- assume that the two stocks are independent, i.e. $\dW_{1,1}(t)\dW_{2,1}(t)=0\cdot\dt$. In such a setup the marginal, cumulative distribution function of each stock can be recovered using Fourier transform (cf.~\cite{OosterleeGrzelakBook}), and the covariance structure can then be imposed using a copula. The basket, $B(T)$, at given maturity $T$ is thus computed via:
\begin{eqnarray*}
 B(T)= F_{S_1(T)}^{-1}(U_1)+F_{S_2(T)}^{-1}(U_2)=S_1(T)+S_2(T),
 \end{eqnarray*}
 where the uniformly distributed random variables $U_1$ and $U_2$ are joined using a Gaussian copula $\mathcal{C}_{\mathcal R}.$ Using such a model we can now price a call option and compute the corresponding implied volatility for a range of copula parameters.

\begin{figure}[h!]
  \centering
    \includegraphics[width=0.45\textwidth]{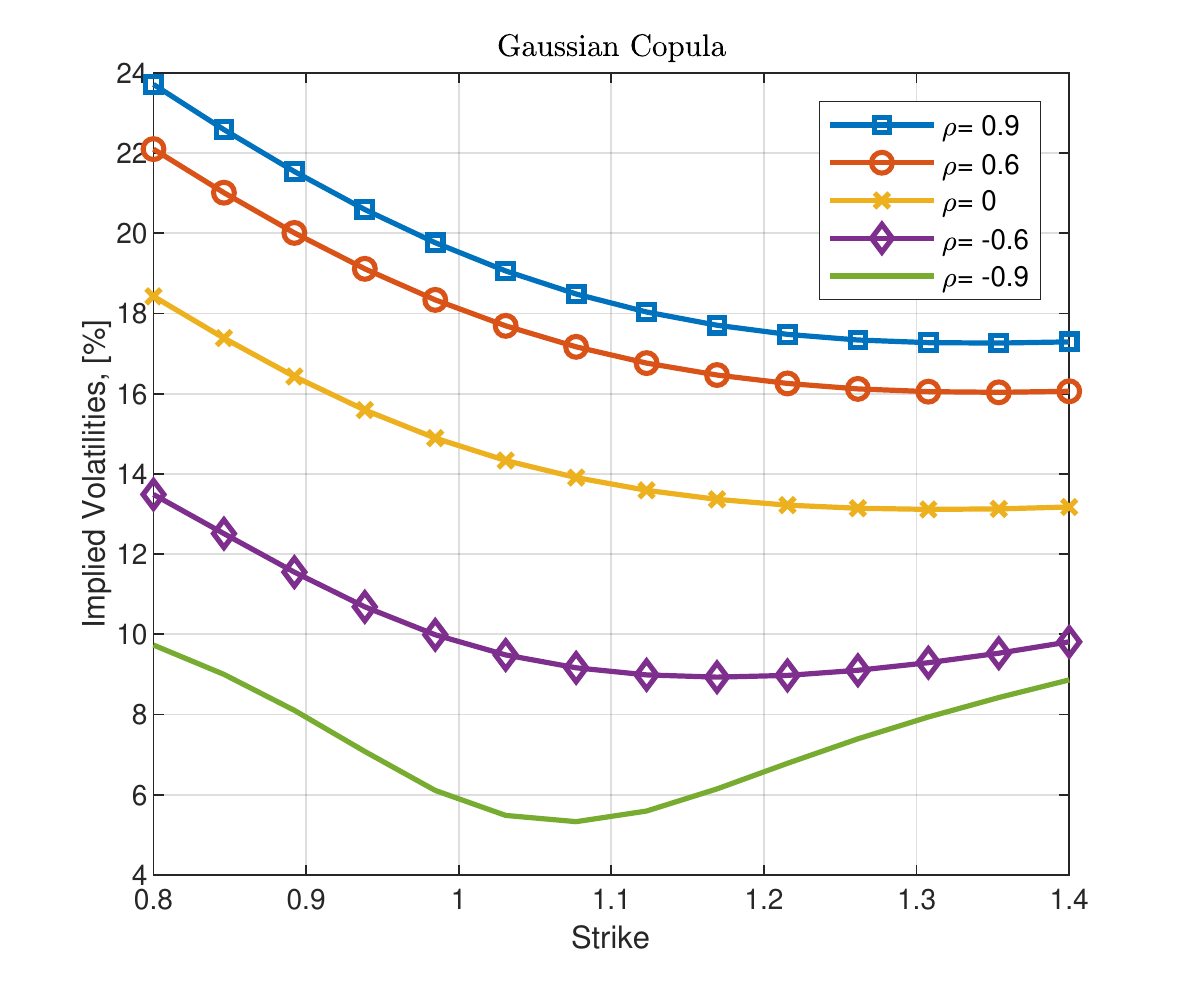}
      \caption{Implied volatilities for a basket $B(T)$ for $d=2$ driven by the Heston model.}
      \label{fig:IVvsCopula}
\end{figure}

 Figure~\ref{fig:IVvsCopula} shows that varying the copula correlation has a significant impact on the level of corresponding implied volatilities. However, it is also clear that while varying the covariance structure alone might help in matching, say ATM index vols, Gaussian copula is unlikely to generate a desired level of skewness -- a problem we already alluded to in the introduction. One potential remedy would be to consider more elaborate copula functions (as e.g. proposed in \cite{Lucic2013}), however we opt for a different approach. As has already been hinted above, and as we shall explain in detail below, choosing Merton's jump diffusion dynamics \cite{Merton:1976} for the synthetic variables mapping individual stocks offers great flexibility in fitting both ATM index vols and skew with excellent precision.

\subsection{Basket Reconstruction via Projection with Collocation}
\label{sec:BasketReconstruction}

As discussed before, our ELV model relies on the availability of the marginal distribution of each asset, $S_i(t)$, at every time $T_j$. Such a distribution may be given explicitly via a specific form of the SDE governing the evolution of $S_i$. Alternatively, the relevant distributions can be inferred from options prices on $S_i$ -- either obtained from the market or generated by some unknown model -- leaving us completely agnostic as to the shape or form of the asset-pricing model. In this section we follow the latter more general approach and derive the distribution through the following well known relation:
\begin{eqnarray}
\label{eqn:densityFromQuotes}
F_{S_i(T_j)}(y)=1+\frac{\partial V_c(t_0,T_j,S_i(t_0),K))}{\partial K}\Big|_{K=y},\;\;\;f_{S_i(T_j)}(y)=\frac{\partial^2 V_c(t_0,T_j,S_i(t_0),K))}{\partial K^2}\Big|_{K=y},
\end{eqnarray}
where $F_{S_i(T_j)}(y)$ is the CDF of $S_i(\cdot)$ observed at time $T_j$ and $f_{S_i(T_j)}(y)$ is the corresponding PDF. We shall take such distributions as given and ultimately try to reconstruct the marginal distributions of the basket $B(T_j)$ defined as the weighted sum of all individual assets. This step will be ultimately performed by ``coupling'' the individual marginals via correleted kernel processes.

However, our first goal is to  project each basket constituent, $S_i(T_j)$, on a polynomial, $g_{i,j}^m(\cdot)$, of synthetic variables $X_i(T_j)$, such as  to ensure a perfect fit to the respective marginal distributions for every set of model parameters $X_i(T_j)$. To this end we resort to the so-called Stochastic Collocation Monte Carlo Sampler (SCMC) discussed in~\cite{Grzelak:2015SCMC} (cf. also an overview in~\ref{sc:scmc}). Using the SCMC method we can express $\log S_i(T_j)$ as:
\begin{equation}
\label{eqn:projection}
    \log S_i(T_j)\approx g^m_{i,j}(X_i(T_j)) = \sum_{k=0}^{m-1}s_{i,j,k}\ell_k(x_{i,j,k},X_i(T_j)),
\end{equation}
where $\ell_k(\cdot)$ are the Lagrange basis function evaluated at the collocation points $x_{i,j,k}$, $k=0,\dots,m-1$ based on the kernel variable $X_i(T_j)$. Alternatively, the representation above can be re-expressed using a polynomial representation,
\begin{equation}
\label{eqn:projection2}
   g^m_{i,j}(X_i(T_j)) = \sum_{k=0}^{m-1}\hat\alpha_{i,j,k} X_i^{k}(T_j) = \hat\alpha_{i,j,0}+\hat\alpha_{i,j,1}X_i(T_j)  + ... + \hat\alpha_{i,j,m-1} X_i^{m-1}(T_j),
\end{equation}
where coefficients $\hat\alpha_{i,j,k}$ are known explicitly~\cite{GrzelakArbitrage:2016}. The expansion in~(\ref{eqn:projection}) ensures that the CDFs of $S_i(T_j)$ and $g^m_{i,j}(X_i(T_j))$ agree on the so-called collocation points, i.e., \[F_{\log S_i(T_j)}(s_{i,j,k})=F_{g_m(X_i(T_j))}(s_{i,j,k}),\;\;\; k=0,\dots,m-1,\]
where $s_{i,j,k}=F_{\log S_i(T_j)}^{-1}(F_X(x_{i,j,k}))$ and where $X_i$ is the collocating variable.

The choice regarding variables $X_i(T_j)$, although in principle free, is crucial for fitting to the marginal distribution of $S_i(T_j)$ and controlling the basket's covariance structure. The most appealing choice regarding variables $X_i$ is a normal distribution~\cite{Grzelak:2015SCMC}, which would imply sampling from multivariate Gaussian copula, thus a computationally cheap, numerical procedure. However, although Gaussian random variables are sufficient for finding optimal $\hat\alpha_{i,j,k}$ in~(\ref{eqn:projection}) guaranteeing excellent fit to marginal distribution of $S_i(T_j)$ such a model will generate low skew of the basket and therefore the final calibration to the index will not be satisfactory (see the remarks in Section \ref{sec:calibration}). As presented later in the article, the non-trivial covariance structure of the underlying assets is necessary for controlling the implied volatility skew. This is confirmed by empirical studies~\cite{Bakshi:2003}.

In the ELV model presented in this article, the kernel that is used is Merton's jump diffusion model \cite{Merton:1976}, which under the risk-neutral measure consists of a Brownian motion and a compound Poisson process, which is defined by:
\begin{eqnarray}
\label{eqn:Merton}
X_i(t)= X_i(t_0)+\left(r-\frac12\sigma_i^2-\xi_{p,i}\E[\e^{J_i}-1]\right)t+\sigma_i W_i(t)+\sum_{k=1}^{X_{\mathcal{P},i}(t)}J_{k,i},
\end{eqnarray}
with $X_i(t_0)=\log S_i(t_0)$, for $\sigma_i>0$, Brownian motion, $W_i(t)$, Poisson process $X_{\mathcal{P},i}(t)$, $t\geq 0$ with parameter $\xi_{p,i}$ such $\E[X_{\mathcal{P},i}(t)]=\xi_{p,i}t$ and where the jump sizes $J_{k,i}$ are i.i.d. given by
$J_{k,i}\sim\mathcal{N}\left(\mu_{J,i},\sigma_{J,i}^2\right)$. For a given index $i$ the jumps, $J_{k,i}$, Brownian motion, $W_i(t)$, and Poisson process, $X_{\mathcal{P},i}(t)$ are assumed to be independent.

\begin{rem}[Alternative dynamics for $X_i$]
Alternatively, one may also consider the more advanced process structure $X_i(\cdot)$ by adding stochastic volatility. Here, however, we will focus on a simplified model, and we will show that such a structure is sufficient for excellent calibration of the market data, even for large baskets consisting of $30$ assets.
\end{rem}


Once the collocating variable has been chosen, we need to establish the mapping procedure from the asset $\log S_i(T_j)$ onto the surrogate variable $X_i(\cdot).$ In order to achieve accurate approximations one needs to firstly determine the so-called collocation points, $x_{i,j}$, $j=1,\dots,N$, of ``surrogate'' variable $X_t(\cdot)$- these collocation points are based on moments of $X_i(t)$. This ensures that the collocation points are the zeros of the orthogonal polynomial corresponding to the distribution of $X_i(\cdot)$, and we can establish the connection with the computation of integrals by Gauss quadrature. Typically, for $N$-collocation points $2N$ moments of $X_i(\cdot)$ are needed. However, for $X_i(\cdot)$ defined in~(\ref{eqn:Merton}) one may, due to the presence of jumps, expect an effect often called ``moment explosion'' which translates to huge moments causing significant numerical instability of the method.
As an alternative approach, we use the collocation points $z_k$, $k=0,\dots,m-1$ corresponding to the standard normal random variable\footnote{The reason why we choose a standard normal distribution in the alternative approach is twofold. First, even for a fundamental distribution as the standard normal results are
highly accurate -- this is also the case in e.g. \cite{GrzelakCLV}. By choosing a different distribution, results may be further enhanced. Secondly,
as mentioned in~\cite{Grzelak:2015SCMC}, choosing the normal distribution is also motivated by the Cameron-Martin Theorem \cite{Cameron1947AnnMath}, which states that polynomial chaos approximations based on the normal distribution converge to any distribution. \label{rem:normal}} $Z\stackrel{\d}{=}\mathcal{N}(0,1)$. Such a strategy implies, for a given time $T_j$, the following relation between $\log S_i(T_j)$, $X_i(\cdot)$ and $Z$:
\begin{eqnarray}
\label{eqn:inversions}
x_{i,j,k}&=&F^{-1}_{X_i(T_j)}(F_Z(z_k)),\\
\log s_{i,j,k}&=&F^{-1}_{\log S_i(T_j)}(F_{X_i(T_j)}(x_{i,j,k})),
\end{eqnarray}
which illustrates how to avoid computation of the collocation points of $X_i(T_j)$ using moments of $X_i(T_j).$
It is important to note that although variable $X_i(T_j)$ is not directly used to compute the collocation points of $S_i(T_j)$ the approximating function $g^m_{i,j}(X_i(T_j))$ is explicitly defined in terms of $X_i$ and points $x_{i,j,k}$ and $\log s_{i,j,k}$. The details on the optimality of such ``substitution'' can be found in~\cite{AntonStoepCollocatingVol}.

\begin{rem}[Grid stretching]
When dealing with heavy-tailed distributions (of leptokurtic type or the distribution is highly skewed), as here proposed for $X_i(T_j)$, to avoid numerically unstable inversions $F^{-1}_{X_i(T_j)}(a)$ for either $a\rightarrow 0$ or $a\rightarrow1$ it is recommended to use the so-called ``grid stretching'' technique~\cite{Grzelak:2015SCMC} which facilitates stable inversions, especially for a high number of the collocation points, $m$.
\end{rem}

Availability of the ChF in closed form allows finding, via Fourier inversion~\cite{OosterleeGrzelakBook}, the corresponding CDF, $F_{X_i(T_j)}(x)$, and thus finding of the corresponding collocation points, $x_{i,j,k}$.
\begin{equation}
\label{eqn:chfMerton}
    \phi_{X_i(t)}(u)=\E\left[\e^{iuX_i(t)}\right]=\exp\left(iu(X_i(t_0)+\bar{\mu}_it)-\frac12\sigma_i^2u^2t+\xi_{p,i}t\e^{iu\mu_{J,i}-\frac12u^2\sigma_{J,i}^2}\right),
\end{equation}
with $\bar\mu_i=r-\frac12\sigma_i^2-\xi_{p,i}(\e^{\mu_{J,i}+\frac12\sigma_{J,i}^2}-1).$

Then, by utilizing the COS method~\cite{OosterleeGrzelakBook}, one is able to determine the corresponding CDF:
\begin{eqnarray}
\label{eqn:CDFCOS}
F_{X_i(T_j)}(y)\approx {\sum_{k=0}^{N_c-1}}\bar{F}_k\psi(a,b,y),\;\;\;\bar{F}_k=\frac{2}{b-a}\Re\left\{\phi_{X_i(T_j)}\left(\frac{k\pi}{b-a}\right)\cdot\exp\left(-i\frac{ka\pi}{b-a}\right)\right\},
\end{eqnarray}
with
\begin{equation}
\psi(a,b,y)=\left\{\begin{array}{ccc}
\frac{b-a}{k\pi}\sin\left(\frac{k\pi(y-a)}{b-a}\right),&\text{for}& k=1,2,\dots,N_c-1,\\
(x-a),&\text{for}& k=0,
\end{array}\right.
\end{equation}
where $N_c$ indicates the number of the expansion terms, $a$ and $b$ are the domain parameters, typically determines in terms of cummulants.


In the next step, we build a multivariate dependence structure between all the marginal distributions using a system of correlated SDEs.
The basket, $B(T_j)$, yields, in terms of the latent variables, the following form:
\begin{eqnarray}
B(T_j)&=&\sum_{i=1}^NS_i(T_j)\approx\sum_{i=1}^N\exp\left(g^m_{i,j}(X_i(T_j))\right)\nonumber\\
&=&\sum_{i=1}^N\exp\left(\sum_{k=0}^{m-1}\hat\alpha_{i,j,k} X_i^{k}(T_j)\right),\label{eqn:mappingApproach}
\end{eqnarray}
with $X_{i}(T_j)$ defined in~(\ref{eqn:Merton}).  Although we have initially assumed that each $X_i$ has its own independent jumps $J_{k,i}$ controlled by $X_{\mathcal{P}_i(t)}$, our numerical studies have shown that without sacrificing much in terms of skew fitting possibilities, one can choose a sparse for of the model with $X_{\mathcal{P}_i(t)}$ being the same for all the assets, i.e., \[ X_{\mathcal{P}_i(t)}\equiv  X_{\mathcal{P}(t)}.\] In such form, the dependence between the individual stocks in the basket is controlled via two main elements: correlated Brownian motions $W_i(T_j)$ and common Poisson process $X_\mathcal{P}(T_j)$ for all Merton's processes, $X_i.$ Thus, the basket is reconstructed using a combination of marginal distributions for each asset. The basket PDF, $f_{B(t)}$, can then be obtained by Monte Carlo. It should be stressed that the sampling from multivariate normal is computationally inexpensive. The expensive part involves the inversion of marginal distributions, but as explained above, we handle this using collocation, mitigating the computational cost quite considerably.

Importantly, the proposed methodology does not rely on a calibration of the process for marginal distributions, but solely on the computation of $\hat\alpha_{i,j,k}$, for $i=1,\dots,N$, $j=1,\dots,N_T,$ $k=1,\dots,m$, and inexpensive samples from the multidimensional distribution of $(X_i(\cdot),\dots,X_N(\cdot))$.

\begin{rem}[Marginal distribution]
We would like to stress that due to the collocation method and the CDF mapping procedure, determined coefficients $\hat\alpha_{i,j,k}$ will ensure agreement between the target CDF, $F_{\log S_i(T_j)}(\cdot)$ and the surrogate, $F_{X_i(T_j)}(\cdot).$ Once the marginal distributions are calibrated for any configurations of model parameters of $X_i(\cdot)$, the smile/skew of the index needs to be matched. Later, in this article, we will show that the model proposed allows for intuitive control of different volatility shapes of the basket, therefore facilitating calibration to the index.
\end{rem}

\section{Basket Calibration to the Index}\label{sect_calibration}
Given that the ELV model, by its construction, guarantees fit to marginal distributions one still needs to calibrate the basket to index volatilities. The brute force strategy of such optimization can be performed using Monte Carlo simulation, i.e., for different model parameters, the samples of $X_i(T_j)$ are fed to basket equation in~(\ref{eqn:mappingApproach}) and the option prices are computed. Such routine is typically sub-optimal as it would require multiple iterations and samplings for $X_i(\cdot).$ To speed up the process and save computational burden, we opt for a different approach, based on moment matching.

Recall that in its most general form the model might appear to have an overwhelming number of parameters to be calibrated. Fortunately, as we have already hinted above, many simplifications can be resorted to without sacrificing calibration precision, while improving tractability and sparsity.

Firstly, we note that the ELV model's heart lies in the idea of construction of the local volatility function; therefore, only the marginal distribution for the basket $B(T_j)$ is relevant. This implies that the calibration can be performed sequentially, or parallel, for every time $T_j$, $j=1,\dots,N_T.$ Secondly, it is in practice safe to assume that the individual stocks share not only the type of the kernel process but also the driving parameters. Specifically, we assume henceforth that all the assets have a common counting process $X_\mathcal{P}(T_j)$, independent i.i.d. jumps $J\equiv J_{k,i}\sim\mathcal{N}(\mu_J,\sigma^2_J)$  and correlated Brownian motions $\dW_i(t) \dW_j(t)=\rho \dt,$ which leads to the following simplified representation:
\begin{eqnarray*}
X_i(T_j)= X_i(t_0)+\big(r-\frac12\sigma^2-\xi_{p}\E[\e^{J}-1]\big)T_j+\sigma W_i(T_j)+\sum_{k=1}^{X_{\mathcal{P}}(T_j)}J_{k,i}.
\end{eqnarray*}
Thus, in order to calibrate the basket $B(T_j)$ to an index we need, for every $T_j$, to determine the set of optimal model parameters $\sigma$, $\xi_p$, $\mu_J$ and $\sigma_J$, driving the model's covariance structure.

Each of the jump parameters has a different effect on the shape of the implied volatility curve, i.e., $\sigma_J$ has a significant impact on the curvature, $\xi_{\tiny p}$ controls the overall level of the implied volatility, whereas $\mu_J$ influences the implied volatility slope (the skew).
\begin{figure}[h!]
  \centering
    \includegraphics[width=4.7cm]{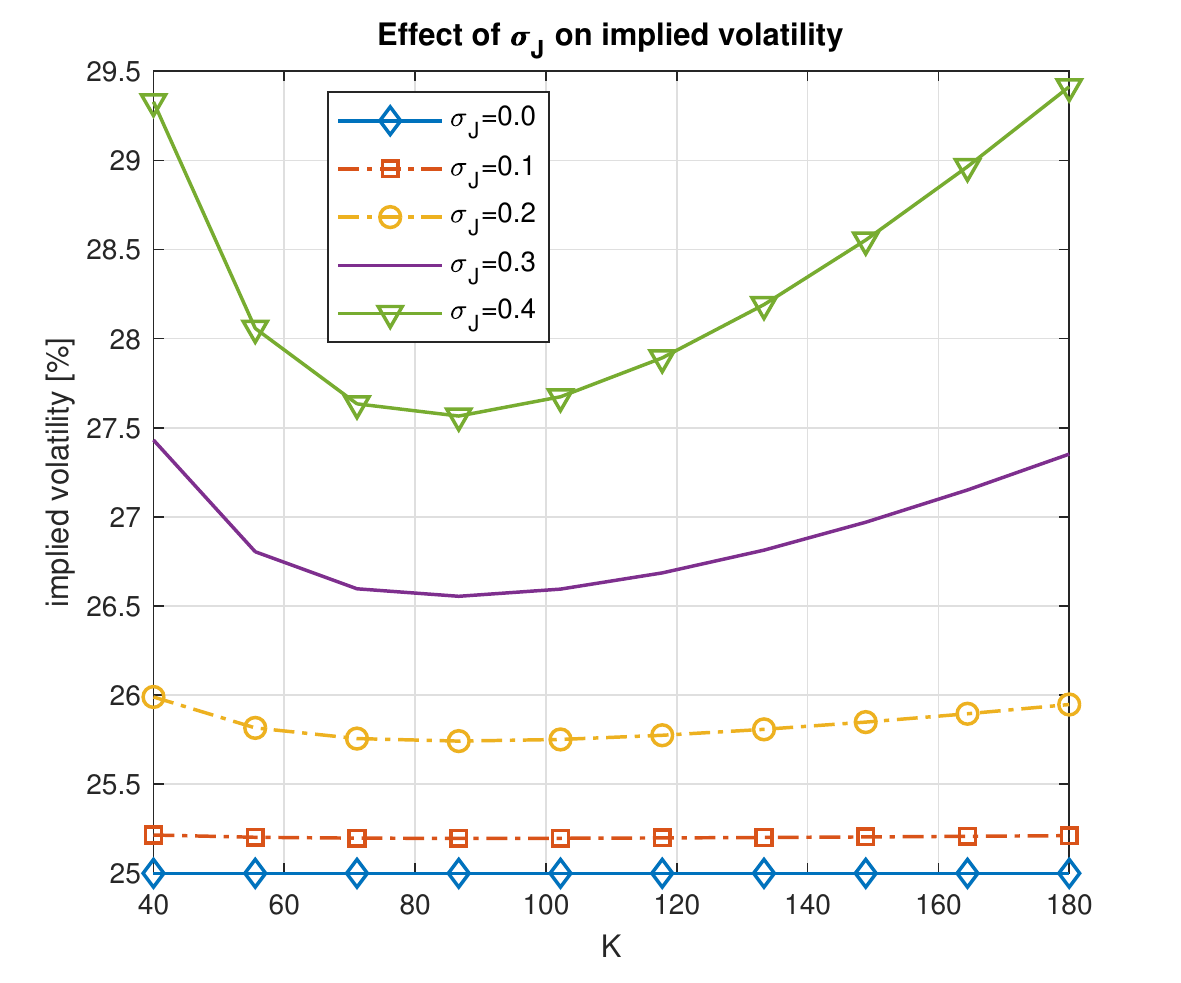}
    \includegraphics[width=4.7cm]{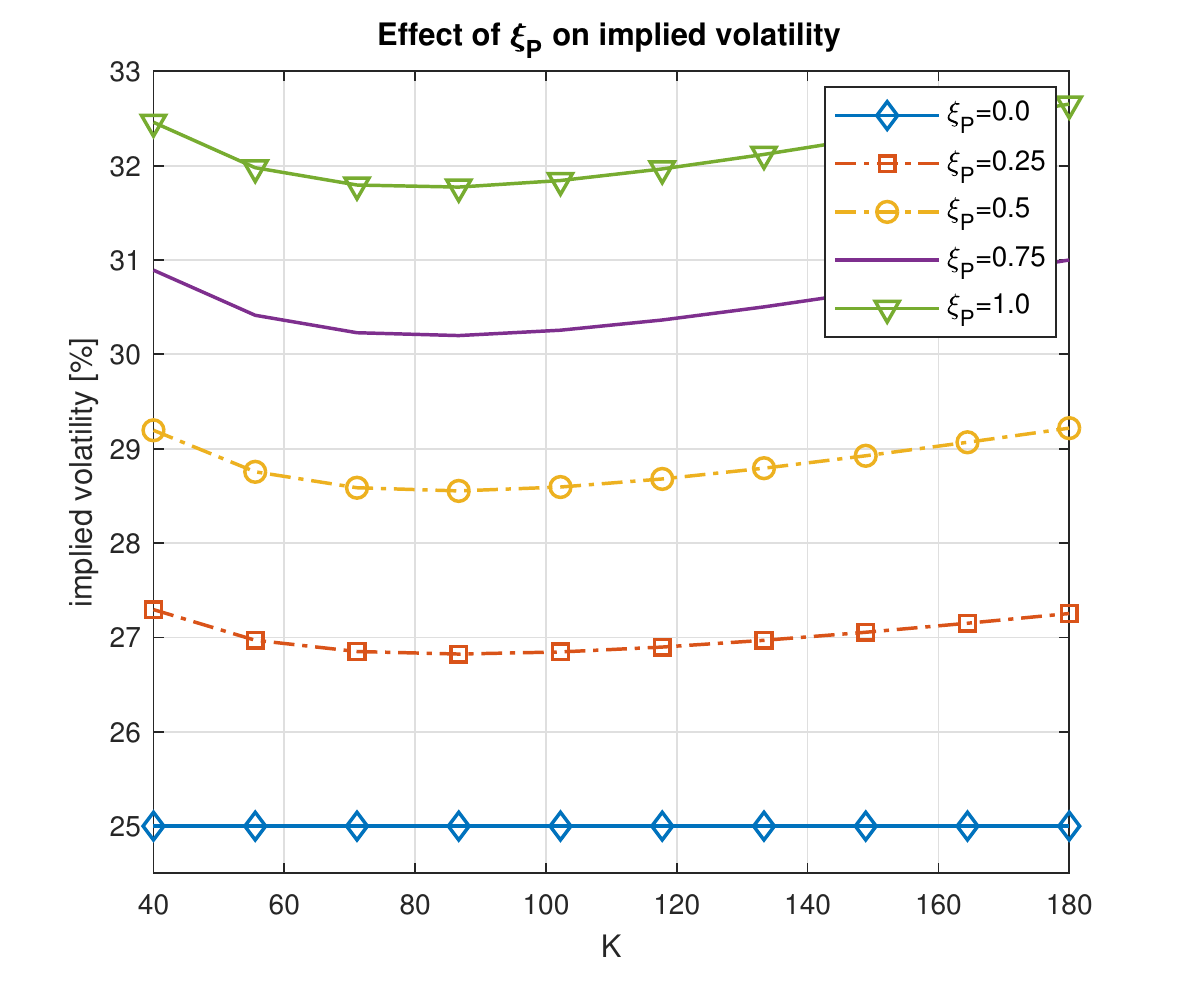}
    \includegraphics[width=4.7cm]{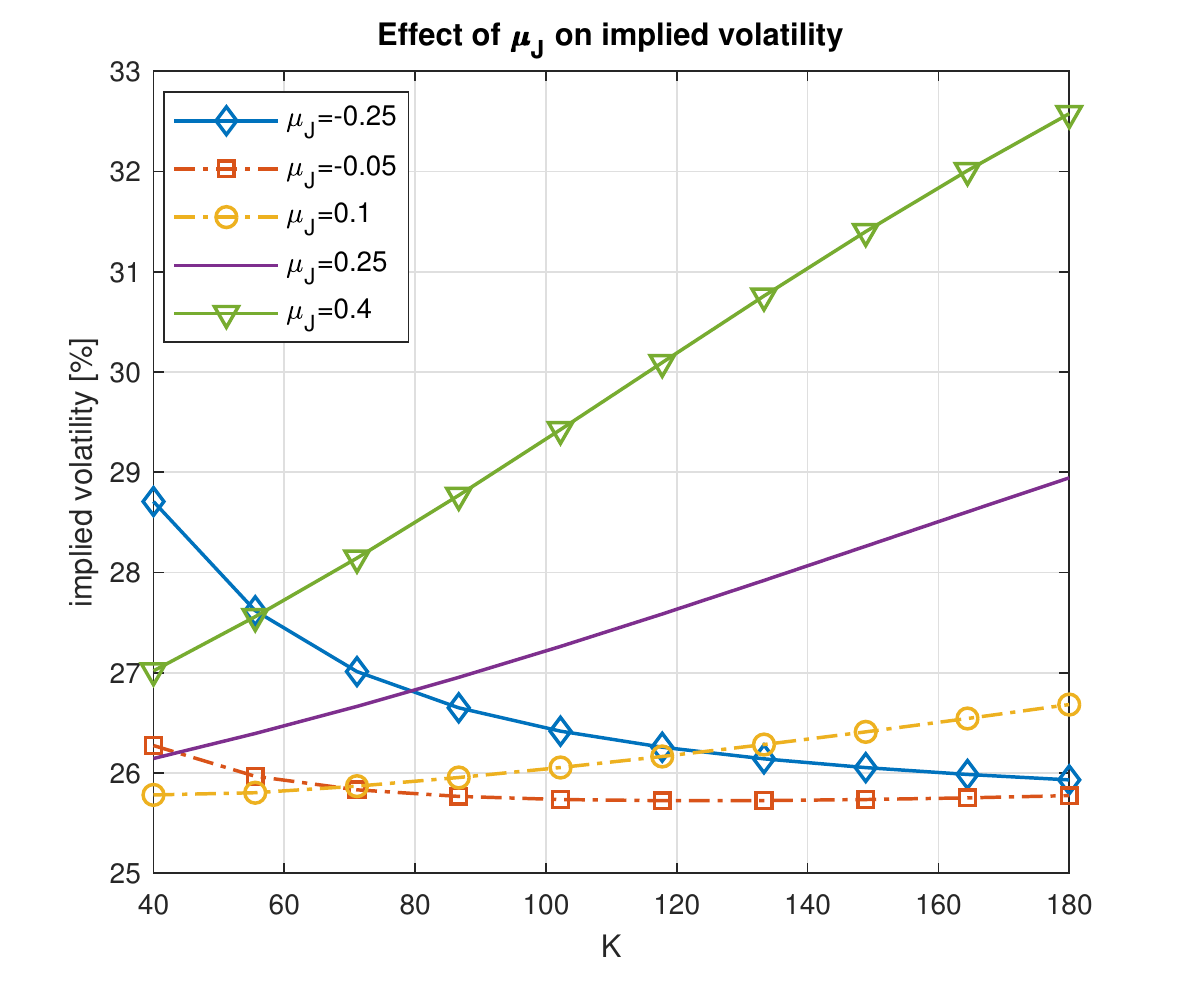}
    \caption{Impact of different jump parameters on the shape of the implied volatility in Merton's jump diffusion model. First figure: impact of $\sigma_J$; Second figure: impact of $\xi_p$; Third figure: impact of $\mu_J.$}
      \label{fig:JumpImpactSingleAsset}
\end{figure}

Having already gained much in terms of tractability, we can further save the computational burden by performing calibration via the moment matching procedure. Given that the density of the index, $f_{I}(y)$, exists and can be calculated from options prices, as per Equation~(\ref{eqn:densityFromQuotes}), the objective is to match moments from the index,$I(t)$, and the basket $B(t)$. It is important to note that the polynomial coefficients $\alpha_{i,j,k}$ are implicitly a function of model parameters, i.e.,
\begin{eqnarray}
\label{eqn:alpha}
\alpha_{i,j,k}:=\alpha_{i,j,k}(\sigma,\xi_p,\mu_J,\sigma_J),
\end{eqnarray}
therefore any change in the model parameters requires re-computation of these coefficients. Since this procedure only relies on the computation of a few inversions of the CDF in Equation~(\ref{eqn:CDFCOS}), it is computationally cheap.

Since the first moment is trivial let us start with the second moment. By definition we have:
\begin{eqnarray*}
\E[B^2(T_j)]&=&\sum_{i_1=1}^N\sum_{i_2=1}^N\omega_{i_1}\omega_{i_2}\E[S_{i_1}(T_j)S_{i_2}(T_j)].
\end{eqnarray*}
One the other hand, assuming that we are able to determine the correlation $\hat\rho(S_{i_1},S_{i_2})$ between $S_{i_1}(T_j)$ and $S_{i_2}(T_j)$ (this correlation will differ from $\rho_{i_1,i_2}\dt=\dW_{i_1}(t)\dW_{i_2}(t)$ between the corresponding Brownian motions), we find:
\[\hat\rho(S_{i_1},S_{i_2})\stackrel{\text{def}}{=}\frac{\E[S_{i_1}(T_j)S_{i_2}(T_j)]-\E[S_{i_1}(T_j)]\E[S_{i_2}(T_j)]}{\sigma_{{i_1}}\sigma_{{i_2}}},\]
where $\sigma_{i,1}$ and $\sigma_{i,2}$ are standard deviations of $S_{i_1}(T_j)$ and $S_{i_2}(T_j)$ respectively (these quantities can be calculated directly using~(\ref{eqn:densityFromQuotes})).
Given that the marginal distributions of $S_{i_1}$ and $S_{i_2}$ are available, both: first moments and the standard deviations can be easily obtained, by for example proper integration of the market-implied densities in~(\ref{eqn:densityFromQuotes}). Then, we have:
\begin{eqnarray}
\label{eqn:ES1S2}
\E[S_{i_1}(T_j)S_{i_2}(T_j)]=\hat\rho(S_{i_1},S_{i_2})\sigma_{{i_1}}\sigma_{{i_2}}+\E[S_{i_1}(T_j)]\E[S_{i_2}(T_j)],\end{eqnarray}
therefore the second moment of the basket becomes:
\begin{eqnarray}
\label{eqn:BasketSecondMoment}
\E[B^2(T_j)]&=&\sum_{i_1=1}^N\sum_{i_2=1}^N\omega_{i_1}\omega_{i_2} \Bigg( \hat\rho(S_{i_1},S_{i_2})\sigma_{{i_1}}\sigma_{{i_2}}+\E[S_{i_1}(T_j)]\E[S_{i_2}(T_j)]\Bigg).
\end{eqnarray}
Since the correlation, $\hat\rho(S_{i_1},S_{i_2})$, measures the linear relationship between assets $S_{i_1}$ and $S_{i_2}$, we will compute this coefficient based on a fewer number of the collocation points~\footnote{The strategy proposed in this part does not require ``re-calibration'' of $\alpha_{i,j,k}$ coefficients, but only neglects the coefficients of higher order.}, $m=2$ which yields the following representation:
\begin{eqnarray}
\label{eqn:corr_i1i2}
\hat\rho(S_{i_1},S_{i_2})&\approx&\rho\left(\e^{\hat\alpha_{i_1,j,0}+\hat\alpha_{i_1,j,1}X_{i_1}(T_j)},\e^{\hat\alpha_{i_2,j,0}+\hat\alpha_{i_1,j,1}X_{i_2}(T_j)}\right)\\
&=&\rho\left(\e^{\hat\alpha_{i_1,j,1}X_{i_1}(T_j)},\e^{\hat\alpha_{i_2,j,1}X_{i_2}(T_j)}\right)=:\bar\rho(S_{i_1},S_{i_2}),\nonumber
\end{eqnarray}
with $X_{i}(\cdot)$ defined in~(\ref{eqn:Merton}), with correlated Brownian motions,
\begin{eqnarray*}
\bar\rho(S_{i_1},S_{i_2})&\stackrel{\text{def}}{=}&\frac{\E[\e^{\hat\alpha_{i_1,j,1}X_{i_1}(T_j)+\hat\alpha_{i_2,j,1}X_{i_2}(T_j)}]-\E[\e^{\hat\alpha_{i_1,j,1}X_{i_1}(T_j)}]\E[\e^{\hat\alpha_{i_2,j,1}X_{i_2}(T_j)}]}{\sqrt{\Var[\e^{\hat\alpha_{i_1,j,1}X_{i_1}(T_j)}]\Var[\e^{\hat\alpha_{i_2,j,1}X_{i_2}(T_j)}]}}.
\end{eqnarray*}
We immediately note the connection between the moments and the corresponding ChF of $X_i(t)$ defined in~(\ref{eqn:chfMerton}) resulting in,
\begin{eqnarray*}
\phi_{X_i(t)}(u)=\E[\e^{iuX_i(t)}],\;\;\;\text{thus}\;\;\;\E[\e^{aX_i(t)}]=\phi_{X_i(t)}(-ia),\;\;a\in\mathbb{R},
\end{eqnarray*}
resulting in the following representation
\begin{eqnarray}
\label{eqn:corr_i1i2Proxy}
\bar\rho(S_{i_1},S_{i_2})=\frac{\E[\e^{c_1X_{i_1}(T_j)+c_2X_{i_2}(T_j)}]-\phi_{X_{i_1}}(-ic_1)\phi_{X_{i_2}}(-ic_2)}{\sqrt{\left(\phi_{X_{i_1}}(-2ic_1)-\phi_{X_{i_1}}^2(-ic_1)\right)\left(\phi_{X_{i_2}}(-2ic_2)-\phi_{X_{i_2}}^2(-ic_2)\right)}},
\end{eqnarray}
with $c_1=\hat\alpha_{i_1,j,1}$, $c_2 = \hat\alpha_{i_2,j,1}$ and $\phi_{X_i(T_j)}$ defined in Equation \ref{eqn:chfMerton}.
The only term that caries the correlation between both processes is the expectation involving correlated Merton's processes. This expectation is given in closed form and it is presented in Lemma~\ref{lem:Merton2d} below.
\begin{lem}[Expectation for a 2D Merton's model]
\label{lem:Merton2d}
For a given 2D Merton's model defined as:
\begin{eqnarray}
\label{eqn:2DMarton}
X_1(t)=\mu_1+\sigma_1 W_1(t)+\sum_{k=1}^{X_{\mathcal{P}}(t)}J_{k,1},\;\;\;X_2(t)=\mu_2+\sigma_2 W_2(t)+\sum_{k=1}^{X_{\mathcal{P}}(t)}J_{k,2},
\end{eqnarray}
with correlated Brownian motions $\d W_1(t)\d W_2(t)=\rho_{1,2}\dt,$ common counting Poisson process $X_{\mathcal{P}}(t)$ with the intensity parameter $\xi_p$ and independent identically distributed jumps, $J\equiv J_{k,\cdot}\sim\mathcal{N}(\mu_J,\sigma_J^2)$, the following expression holds:
\begin{eqnarray}
\nonumber
\E[\e^{aX_{1}(T)+bX_{2}(T)}|\F(t_0)]&=&\exp\left[a\mu_1+b\mu_2+\frac12\hat\sigma^2T+\xi_pT\left(\e^{(a+b)\mu_J+\frac12(a^2+b^2)\sigma_J^2}-1\right)\right]\\
&=:&\omega_X(a,b),
\label{eqn:Expectation2D}
\end{eqnarray}
where $\hat\sigma^2=a^2\sigma_1^2+b^2\sigma_2^2+2\rho_{1,2}ab\sigma_1\sigma_2$.
\begin{proof}
Proof can be found in~\ref{app:proofMerton2D}.
\end{proof}
\end{lem}
Utilizing the results above we find the closed form solution for the expectation and therefore the correlation coefficient, $\bar\rho(S_{i_1},S_{i_2})$,
by simply setting where $c_1 = \hat\alpha_{i_1,j,1}$, $c_2=\hat\alpha_{i_2,j,1}$, $\mu_i=X_i(t_0)+\big(r-\frac12\sigma^2-\xi_{p}\E[\e^{J}-1]\big)T$, and $\sigma_1=\sigma_2=\sigma.$

Once the correlation coefficient, $\bar\rho(S_{i_1},S_{i_2})$, is determined one is able to perform the ATM calibration of the basket to the index.
\begin{eqnarray}
\label{eqn:opitmization1}
\min_{\sigma_J,\mu_J,\xi_p,\rho}\left(\E[B^2(T_j)]-\E[I^2(T_j)]\right)^2,\;\;\;\E[I^2(T_j)]=\int_\R x^2f_{I(T_j)}(x)\dx,
\end{eqnarray}
where the index density $f_{I(T_j)}(x)$ is implied from the option quotes for the index~(\ref{eqn:densityFromQuotes}) and $\E[B^2(T_j)]$ is defined in~(\ref{eqn:BasketSecondMoment}). It is important to note that, each change of the model parameters of the kernel process does impact parameters $\alpha_{i_1,j,1}$ and $\alpha_{i_2,j,1}$, implying that the mapping coefficients, $\alpha_{\cdot}$, need to be recomputed at each iteration. This however is extremely cheap operation as it only requires the CDF mappings presented in~(\ref{eqn:inversions}).

Following a similar strategy a third moment for the basket can be derived. The detailed derivations for the approximation results are presented in~\ref{appendix:thirdmoment}.
By definition of the third moment we have:
\begin{eqnarray*}
\E[B^3]&=&\sum_{i_1=1}^N\sum_{i_2=1}^N\sum_{i_3=1}^N\omega_{i_1,i_2,i_3}\Big[\hat\rho({S_{i_1,i_2},S_{i_3}})\sigma_{i_1,i_2}\sigma_{{i_3}}+\E[S_{i_1,i_2}]\E[S_{i_3}]\Big]\\
&=&\sum_{i_1=1}^N\sum_{i_2=1}^N\sum_{i_3=1}^N\omega_{i_1,i_2,i_3}\left[\hat\rho({S_{i_1,i_2},S_{i_3}})\sigma_{{i_1,i_2}}\sigma_{{i_3}}+\Big(\hat\rho(S_{i_1},S_{i_2})\sigma_{{i_1}}\sigma_{{i_2}}+\E[S_{i_1}]\E[S_{i_2}]\Big)\E[S_{i_3}]\right],
\end{eqnarray*}
with $\omega_{i_1,i_2,i_3}:=\omega_{i_1}\omega_{i_2}\omega_{i_3}$, and $S_{i_1,i_2}:=S_{i_1}S_{i_2}$. The first moments $\E[S_{i_1}]$, $\E[S_{i_2}]$ and $\E[S_{i_3}]$ are known explicitly using the market market data for individual assets. Standard deviations, $\sigma_{i_1}$ and $\sigma_{i_2}$, of assets $S_{i_1}(T_j)$ and $S_{i_2}(T_j)$ can be computed utilizing the market implied density in~(\ref{eqn:densityFromQuotes}).
Correlation, $\hat\rho({S_{i_1,i_2},S_{i_3}})$, is approximated, as for the second moment, and yields:
\begin{eqnarray*}
\hat\rho({S_{i_1,i_2},S_{i_3}})\approx\frac{\omega_X(c_1,c_2,c_3)-\omega_X(c_1,c_2)\phi_{X_{i_3}}(-ic_3)}{\sqrt{\left(\omega_X(2c_1,2c_2)-\omega_X^2(c_1,c_2)\right)\left(\phi_{X_{i_3}}(-2ic_3)-\phi_{X_{i_3}}^2(-ic_3)\right)}},
\end{eqnarray*}
where:  $c_1:=\hat\alpha_{i_1,j,1}$, $c_2:=\hat\alpha_{i_2,j,1}$, $c_3:=\hat\alpha_{i_3,j,1}$ with $\omega_X(c_1,c_2,c_3)$ defined in~(\ref{eqn:omega3D}) and $\omega_X(c_1,c_2)$ in~(\ref{eqn:Expectation2D}). Finally, the variance $\sigma_{{i_1,i_2}}^2$ is given explicitly by:
\begin{eqnarray}
\nonumber
\sigma_{{i_1,i_2}}^2&:=&\Var[S_{i_1}S_{i_2}]=\E[S_{i_1}S_{i_2}] - \E[S_{i_1}]\E[S_{i_2}]\\
\label{eqn:sigma2_3D}
&=&\hat\rho(S_{i_1},S_{i_2})\sigma_{{i_1}(T_j)}\sigma_{{i_2}(T_j)},
\end{eqnarray}
where $\hat\rho(S_{i_1},S_{i_2})$ is given in~(\ref{eqn:corr_i1i2}) and it is approximated by $\bar\rho(S_{i_1},S_{i_2})$ and derived in~(\ref{eqn:corr_i1i2Proxy}). Availability of the third moment allows for an extended optimization problem proposed in~(\ref{eqn:opitmization1}), i.e.,
\begin{eqnarray}
\label{eqn:opitmization2}
\min_{\sigma_J,\mu_J,\xi_p,\rho}\sum_{k=2}^3\left(\E[B^k(T_j)]-\E[I^k(T_j)]\right)^2,\;\;\;\E[I^k(T_j)]=\int_\R x^kf_{I(T_j)}(x)\dx,
\end{eqnarray}
which will not only allow for the ATM fit but also the level of the implied volatility skew can be matched.

The quality of the approximating moment formulae will be discussed in the follow-up section (see Table~\ref{tbl:Moments}), where a basket consisting of 5 assets will be considered.

\section{Numerical Results}\label{sect_numerical}
This numerical section is dedicated to the numerical aspects and implementation details of the proposed model. We start with a step-by-step example outlining the procedure of constructing a 5-dimensional basket including the calibration of the collocation method to market-implied volatilities. We follow up with a discussion of the impact of model parameters on the resulting basket implied volatility and a finally conclude with a fully-fledged calibration of a high-dimensional basket to market data.

\subsection{Illustrative 5D Example}
\label{sec:5Dexample}
This section presents a step-by-step procedure for building up a basket using the ELV method. We illustrate in detail the complete process, starting from individual asset calibration to basket construction and the impact of the model parameters on the basket. In this example, we consider a basket consisting of 5 assets~\footnote{1) UnitedHealth; 2) Home Depot; 3) Goldman Sachs; 4) Microsoft Corp; 5) salesforce.com Inc} that are a part of the DJIA 30. In the experiment, we take $T_j=1y$ as of 8/12/2021. The implied volatilities for this set of assets are presented in Table~\ref{tbl:impliedVolsMarket}.
\begin{table}[!h]
\caption{Implied volatilities for 5 assets, for $T=1$, observed on 8/12/2021. Spot values for the underlying assets are: $S_1(t_0)=468.86$, $S_2(t_0)=411.25$, $S_3(t_0)=397.32$, $S_3(t_0)=334.97$, $S_5(t_0)=266.31$}
\label{tbl:impliedVolsMarket}
\centering\footnotesize
\begin{tabular}{c|c|c|c|c|c|c|c|c|c|c|c}
 \hline
asset &0.8&	0.85&	0.9&	0.95&	0.975&	1	&1.025&	1.05&	1.1&	1.15&	1.2\\\hline\hline
stock 1&32.20&31.00&29.98&29.07&28.65&28.25&27.87&27.51&26.85&26.28&25.83\\
stock 2&29.96&28.87&27.99&27.31&27.03&26.79&26.58&26.41&26.15&25.98&25.90\\
stock 3&32.23&31.07&30.11&29.34&29.02&28.73&28.48&28.26&27.90&27.65&27.47\\
stock 4&31.86&30.56&29.39&28.35&27.88&27.43&27.02&26.64&25.99&25.47&25.09\\
stock 5&34.58&33.84&33.25&32.76&32.55&32.35&32.17&32.00&31.71&31.46&31.26\\\hline
\end{tabular}
\end{table}

As the first step of our modelling procedure, we parametrize the market-implied volatilities using the SABR-based parametrization formula~\cite{Hagan:2002}. The parameters obtained in this step are presented in Table~\ref{tbl:haganCalibration}.

\begin{table}[!h]
\caption{Implied volatility parametrization using parameterization in~\cite{Hagan:2002}. The meaning of the parameters is as follows: $\beta_H$ -``exponent beta'', $\alpha_H$- ``initial vol'', $\rho_H$- ``correlation'' and $\gamma_H$- ``vol-vol''. In the calibration procedure $\beta_H$ was fixed at 0.9. Interest rates were fixed at $r=0.0.$}
\label{tbl:haganCalibration}
\centering\footnotesize
\begin{tabular}{c|c|c|c|c}
 \hline
asset &$\beta_H$&	$\alpha_H$&	$\rho_H$&	$\gamma_H$\\\hline\hline
stock 1&0.90&0.52&         -0.52&0.55\\
stock 2&0.90&0.48&         -0.25&0.65\\
stock 3&0.90&0.51&         -0.30&0.66\\
stock 4&0.90&0.49&         -0.48&0.66\\
stock 5&0.90&0.56&         -0.27&0.46\\\hline
\end{tabular}
\end{table}

Once each implied volatility skew/smile is parameterized the next step is use~(\ref{eqn:densityFromQuotes}) and using the collocation method project the so-called ``market distribution'' on $X_i$. We set the number of the collocation points to $m=7$, and as discussed in Remark in Section~\ref{sec:BasketReconstruction} we use the grid stretching technique to control the tail behaviour. Firstly, we take $Z~\sim\mathcal{N}(0,1)$ and we set the corresponding min/max quantiles at $q_1=0.01$ and $q_{m=7}=0.99$. Once the corresponding collocation points $z_k$, $k=1,\dots,m$ are determined~\cite{Grzelak:2015SCMC} the inverse for~$x_{i,k}$ in~(\ref{eqn:inversions}) need to take place. Note that in principle each asset, $S_i(T_j)$, may be projected on the same variable $X_i.$ Even if $X_i$'s are correlated (via Brownian motion) it does not affect marginal distributions for $S_i$, but it will be crucial in the basket construction. Table~\ref{tbl:Collocation} illustrates the details on computation of triples $(z_{i,k},x_{i,k},s_{i,k})\equiv(z_{k},x_{k},s_{i,k})$, $k=1,\dots,7,$ $i=1,\dots,5.$

\begin{table}[!h]
\caption{Projection details for the collocation method. Parameters for $X_i$ were the following: $\sigma=0.1$, $\sigma_J=0.02$, $\mu_J=0.0$, $\xi_p=0.25$ and $T=0.5$. Note that $T$ here is considered as a parameter therefore it can be fixed for all the assets. In the experiment we have also taken the stretched grid with $p=0.01$, as proposed in~\cite{Grzelak:2015SCMC}, implying the the lower and upper bound for the CDF, $F_Z(z_1)$ and $F_Z(z_7)$. } \label{tbl:Collocation}
\centering\footnotesize
\begin{tabular}{c|c|c|c|c|c|c|c}
 &$k=1$&$k=2$&$k=3$&$k=4$&$k=5$&$k=6$&$k=7$\\\hline
$z_k$ &-3.7504&-2.3668&-1.1544&0&1.154&2.3668&3.7504\\\
$F_Z(z_k)$&0.0100&0.0710&0.237&0.5000&0.7630&0.929&0.9900\\
$x_k$& -0.1684&   -0.1074&   -0.0539&   -0.0030&    0.0478&    0.1013&   0.1623\\\hline
$s_{1,k}$&4.9851&5.6027&5.9341&6.1545&6.3270&6.4891&6.6839\\
$s_{2,k}$&4.9591&5.5291&5.8174&6.0101&6.1734&6.3512&6.6100\\
$s_{3,k}$&4.7692&5.4420&5.7672&5.9771&6.1500&6.3350&6.6025\\
$s_{4,k}$&4.5586&5.2737&5.6126&5.8221&5.9828&6.1404&6.3508\\
$s_{5,k}$&4.4752&4.9965&5.3177&5.5592&5.7702&5.9855&6.2538\\
\end{tabular}
\end{table}

Given the the projection points presented in Table \ref{tbl:Collocation} and the underlying process $X_i(\cdot)$ we are able to replicate the marginal distribution functions of the assets observable in the market. By simulating a process $X_i(T_j)$ for every asset $i=1,\dots,N$ with the parameters specified in Table~\ref{tbl:Collocation} we evaluate the approximating polynomial for $m=7$ with $T=1y,$
\begin{equation}
\label{eqn:individualAsset}
    \log S_i(T)\approx g^7_{i}(X_i(T)) = \sum_{k=0}^{m-1}s_{i,k}\ell_k(x_{k},X_i(T)),\;\;\;\ell_k(x_k,X_i(T))=\prod_{l=1,\\l\neq
k}^m\frac{X_i(T)-x_l}{x_k-x_l},
\end{equation}
with points $x_{k}$ and $s_{i,k}$ defined in Table~\ref{tbl:Collocation} and $X_i(\cdot)$ defined in~(\ref{eqn:Merton}).
Ultimately, all the kernel processes $X_i$, $X_j$ need to be correlated with a given correlation coefficient. This, however, is not necessary if we are only interested in marginal distributions, thus in a fit of each asset to market implied volatility.

The calibration quality for the considered five assets is presented in Table~\ref{tbl:IV_calibration}. Again, we report excellent results; the error does not exceed $0.2\%$.

\begin{table}[!h]
\caption{Table illustrates the calibration of the collocation method to market-implied volatilities. ``mkt'' stands for the market implied volatility obtained from the parameterization of implied volatility, ``coll'' represents the results obtained from the collocation method and ``err'' is the difference of implied volatilities, in $\%$. } \label{tbl:IV_calibration}
\centering\footnotesize
\begin{tabular}{c|c|c|c|c|c|c|c|c|c|c|c|c}
 \multicolumn{2}{c}{}&\multicolumn{9}{c}{relative strike}\\ \hline
asset&IV \% &0.8&	0.85&	0.9&	0.95&	0.975&	1	&1.025&	1.05&	1.1&	1.15&	1.2\\\hline\hline
$S_1$&mkt&32.2&31.0&30.0&29.1&28.7&28.3&27.9&27.5&26.9&26.3&25.8\\
&coll&32.1&31.0&30.0&29.0&28.6&28.2&27.8&27.4&26.8&26.2&25.7\\\cline{2-13}
&err&0.1&0.0&0.0&0.1&0.1&0.1&0.1&0.1&0.1&0.1&0.1\\\hline\hline
$S_2$&mkt&29.9&28.9&28.0&27.3&27.0&26.8&26.6&26.4&26.1&26.0&25.9\\
&coll&29.9&28.9&28.0&27.3&27.0&26.8&26.5&26.4&26.1&25.9&25.9\\\cline{2-13}
&err&0.0&0.0&0.0&0.0&0.0&0.0&0.0&0.0&0.0&0.0&0.1\\\hline\hline
$S_3$&mkt&32.2&31.1&30.2&29.4&29.0&28.7&28.5&28.2&27.9&27.6&27.5\\
&coll&32.0&31.0&30.0&29.3&28.9&28.7&28.4&28.2&27.8&27.6&27.4\\\cline{2-13}
&err&0.2&0.1&0.1&0.1&0.1&0.1&0.1&0.1&0.1&0.1&0.1\\\hline\hline
$S_4$&mkt&31.9&30.6&29.4&28.4&27.9&27.4&27.0&26.6&26.0&25.5&25.1\\
&coll&31.7&30.4&29.2&28.2&27.7&27.3&26.9&26.5&25.9&25.4&25.0\\\cline{2-13}
&err&0.2&0.2&0.2&0.1&0.1&0.1&0.1&0.1&0.1&0.1&0.1\\\hline\hline
$S_5$&mkt&34.5&33.9&33.3&32.8&32.6&32.4&32.2&32.0&31.7&31.5&31.3\\
&coll&34.6&33.9&33.3&32.8&32.6&32.4&32.2&32.0&31.7&31.5&31.3\\
&err&-0.1&0.0&0.0&0.0&0.0&0.0&0.0&0.0&0.0&0.0&0.0\\\hline
\end{tabular}
\end{table}

As the final step of this illustrative example we construct a basket, $B(T)=\sum_{i=1}^5S_i(T)$ with $S_i(t)$ given in~(\ref{eqn:individualAsset}). For this purpose, we utilize Equation~(\ref{eqn:mappingApproach}), where the processes $X_i$ need to be simulated with Monte Carlo. As discussed before, we consider a standard Poisson process, $X_{\mathcal{P}}(T)$, for all the underlying assets and correlated Brownian Motions, $W(t).$ These processes can be pre-simulated and then fed into the asset Equation~(\ref{eqn:individualAsset}). This procedure will produce Monte Carlo paths for the basket, $B(t)$.

The next step is to determine optimal kernel model parameters. The objective is to find the parameters that will generate basket implied volatility as close as possible to the implied volatility of the index. In this illustrative experiment, we do not consider an index; however, we can analyze the quality of the approximating formulae and the impact of the model parameters on the basket volatilities.

In the calibration procedure, we will follow Equation~(\ref{eqn:opitmization2}), which relies on fitting based on the variance and the skew. In Table~\ref{tbl:Moments} the numerical results for estimated variance and the skew for varying model parameters are shown. The results are excellent, especially for the standard deviation, $\sigma_B$, where the error is about $0.1$.

\begin{table}[!h]
\caption{Table illustrates the quality of the approximation of the moments derived for a considered basket $B(T)$, $T=1$, and varying set of the model parameters. The first 5 columns of the table illustrate particular parameter configuration (dots indicate unchanged parameters).  $\sigma_B:=\sqrt{\Var[B(T)]}$. } \label{tbl:Moments}
\centering\footnotesize
\begin{tabular}{c|c|c|c|c||c|c||c|c}
\multicolumn{5}{c||}{parameters}&\multicolumn{2}{c||}{Monte Carlo}&\multicolumn{2}{c}{Approximation}\\ \hline
$\sigma$ & $\sigma_J$& $\mu_J$& $\xi_p$& $\rho$& $\sigma_{B}$& $\E[B^3]/\sigma_B^3$& $\sigma_{B}$&$\E[B^3]/\sigma_B^3$\\\hline\hline
0.2&0.02&0.0&0.25&0.5&28.17&	98.42&28.1&	90.84\\\hline\hline
0.2&$\cdot$&$\cdot$&$\cdot$&$\cdot$&28.19&	98.17&28.17&	90.17\\\hline
0.5&$\cdot$&$\cdot$&$\cdot$&$\cdot$&28.18&	98.25&28.19&	89.95\\\hline\hline
$\cdot$&0.01&$\cdot$&$\cdot$&$\cdot$&28.19&	98.21&28.18&	90.02\\\hline
$\cdot$&0.04&$\cdot$&$\cdot$&$\cdot$&27.88&	101.25&27.68&	94.85\\\hline\hline
$\cdot$&$\cdot$&-0.1&$\cdot$&$\cdot$&29.77&	84.48&29.61&	78.13\\\hline
$\cdot$&$\cdot$&-0.25&$\cdot$&$\cdot$&31.18&	74.23&31.86&	63.47\\\hline\hline
$\cdot$&$\cdot$&$\cdot$&0.1&$\cdot$&28.2	&98.11&28.17&	90.13\\\hline
$\cdot$&$\cdot$&$\cdot$&0.15&$\cdot$&28.19&	98.22&28.14&	90.47\\\hline\hline
$\cdot$&$\cdot$&$\cdot$&$\cdot$&0.1&19.44&	277.62&19.33&	269.78\\\hline
$\cdot$&$\cdot$&$\cdot$&$\cdot$&0.8&33.31&	62.87&33.31&	56.03\\\hline
\end{tabular}
\end{table}

In the final part of this section, we illustrate, in Table~\ref{tbl:timing}, the timing results. As presented in the table, the most time-consuming part is to parameterize market discrete implied volatilities into a parametric form. Typically, such a process is performed only once, and only the relevant, optimized parameters are stored. The collocation method's calibration and the collocation points' generation is a considerably cheap operation. Finally, the Monte Carlo simulation required sampling of all the underlying processes and construction of the basket.

\begin{table}[!h]
\caption{Timing results in seconds for the basket consisting of 5 assets. The results were performed on a standard home PC where no parallelization was used. Timings are aggregated for all the assets and are reported in seconds. ``IV param'' indicates the time needed to parameterize the implied volatilities- this process may be considered irrelevant for this framework as it is a part of data processing. ``Gen.$x_i$'' indicates the time needed to generate the collocation points $z_i$ and $x_i$ based on the process $X_i$. ``Gen.$y_i$'' represents the time needed to calibrate the collocation method via the inverse of the corresponding CDFs. ``Monte Carlo'' corresponds to the complete basket simulation with five assets and 50k paths for each stock. }
\label{tbl:timing}
\centering\footnotesize
\begin{tabular}{c|c|c|c}
 \hline
 IV param. (s) &	Gen.$x_i$ (s) &	Gen.$y_i$ (S) &Monte Carlo (s)\\\hline\hline
0.5765&0.040    &0.0876&0.240\\ \hline
\end{tabular}
\end{table}

In the next section the impact of the model parameters on basket implied volatility will be studied.

\subsection{Impact of the Model Parameters on Basket Implied Volatility}
This section analyses the impact of the kernel model parameters on the basket implied volatilities induced by the ELV model. As described earlier, due to the collocation method, a particular choice of model parameters does not have a material impact on the fit of individual assets- they, due to the mapping procedure, remain intact. However, it has an impact on the covariance structure of the basket. This effect can be measured in terms of implied volatilities. In the experiment we set the parameters of process $X(t)$ (all the parameter values are the same for all the assets) $\sigma = 0.10$, $\sigma_J= 0.02$, $\mu_J=0$, $\xi_p= 0.25$, $T = 0.5$ with the correlation between all the Brownian motions, $\dW_i(t)\dW_j(t)=\rho\dt$, $\rho=0.5$  and measure the impact on the implied volatility of the basket, $B(T)$- the basket consists of 5 assets and it's construction is defined in Section~\ref{sec:5Dexample}.

\begin{figure}[h!]
  \centering
    \includegraphics[width=0.48\textwidth]{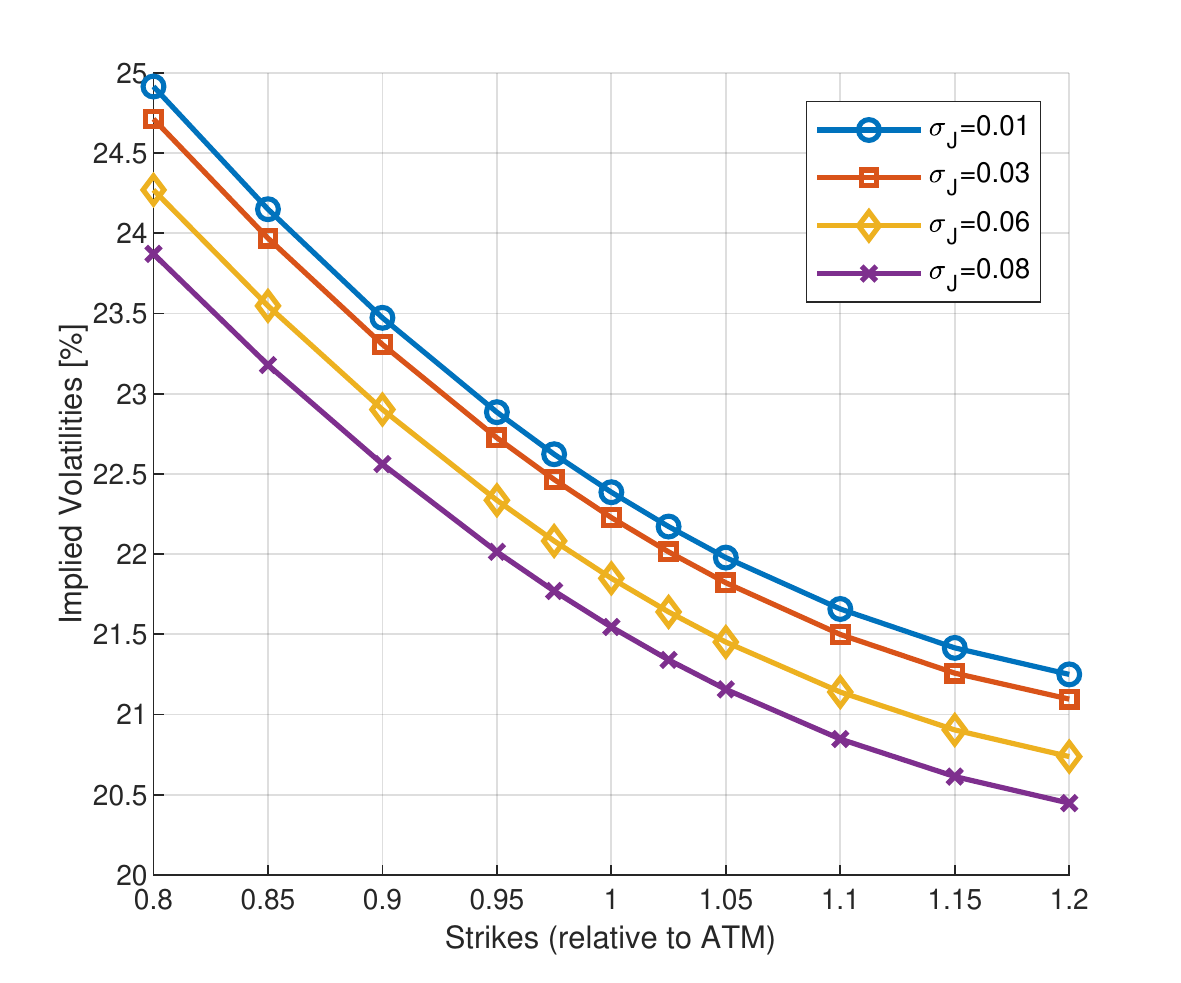}
    \includegraphics[width=0.48\textwidth]{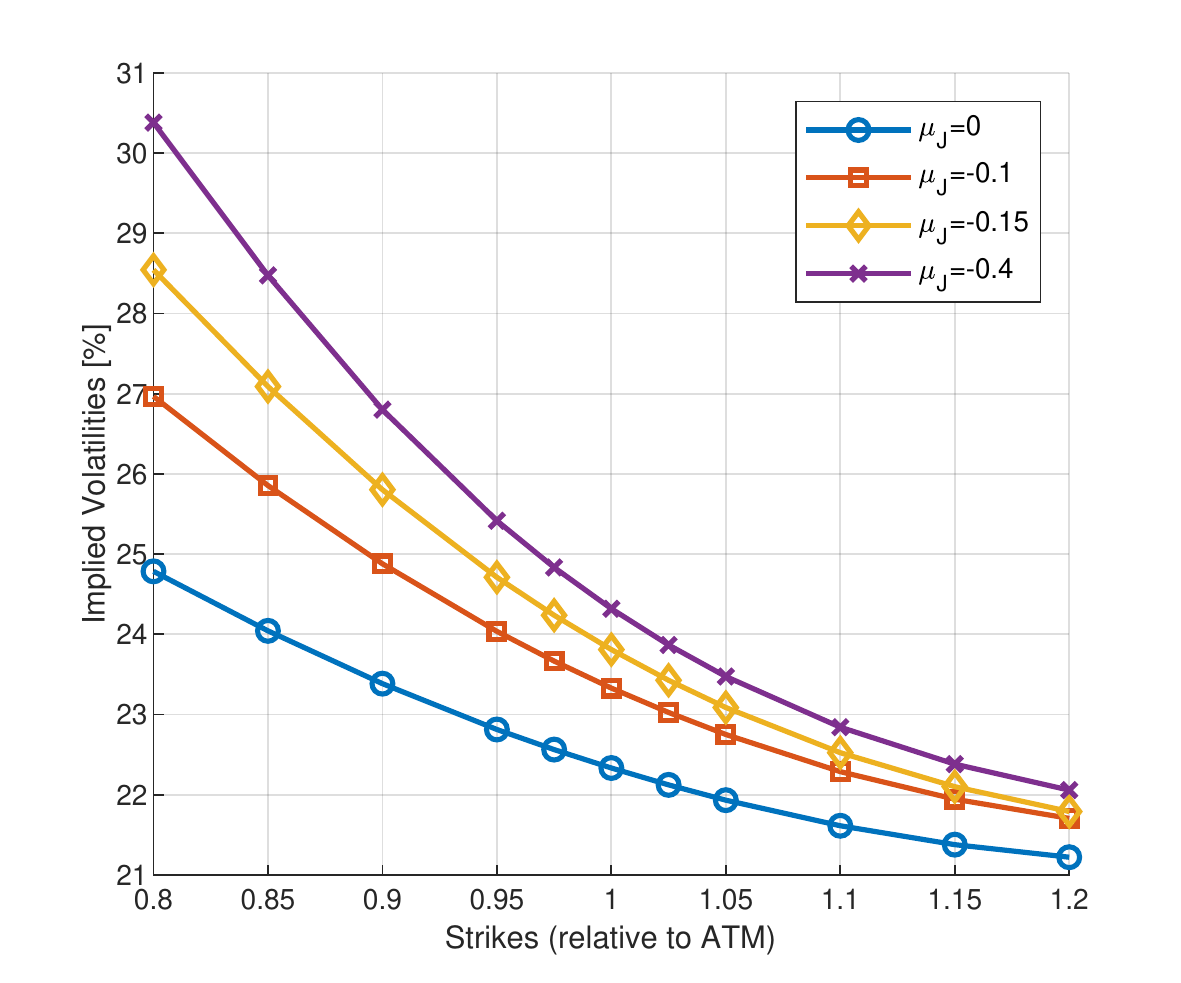}
    \caption{Impact of different jump parameters on the shape of the basket implied volatility in the ELV model. First figure: impact of $\sigma_J$; Second figure: impact of $\mu_J$.}
      \label{fig:impact1}	
\end{figure}

\begin{figure}[h!]
  \centering
    \includegraphics[width=0.48\textwidth]{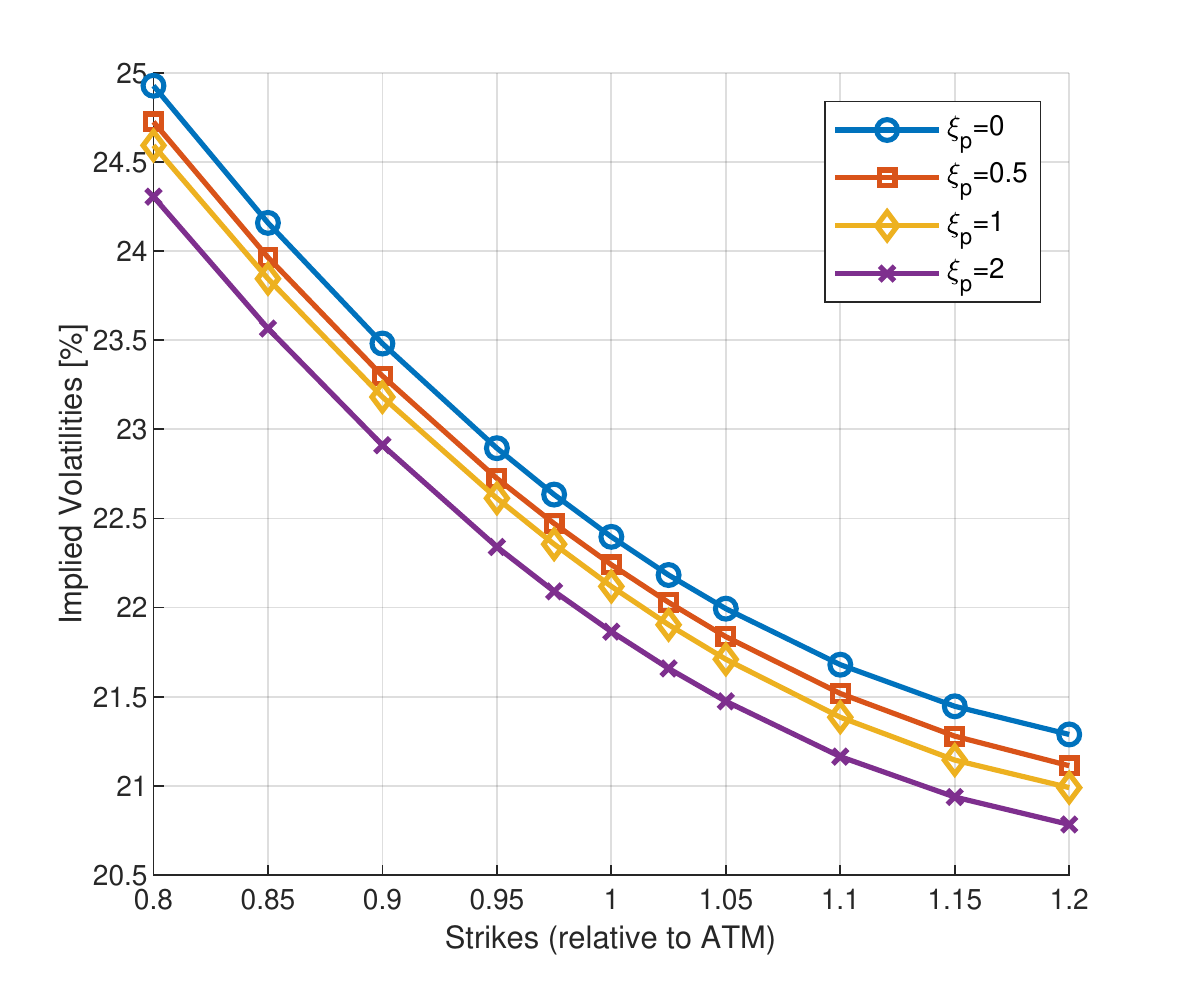}
    \includegraphics[width=0.48\textwidth]{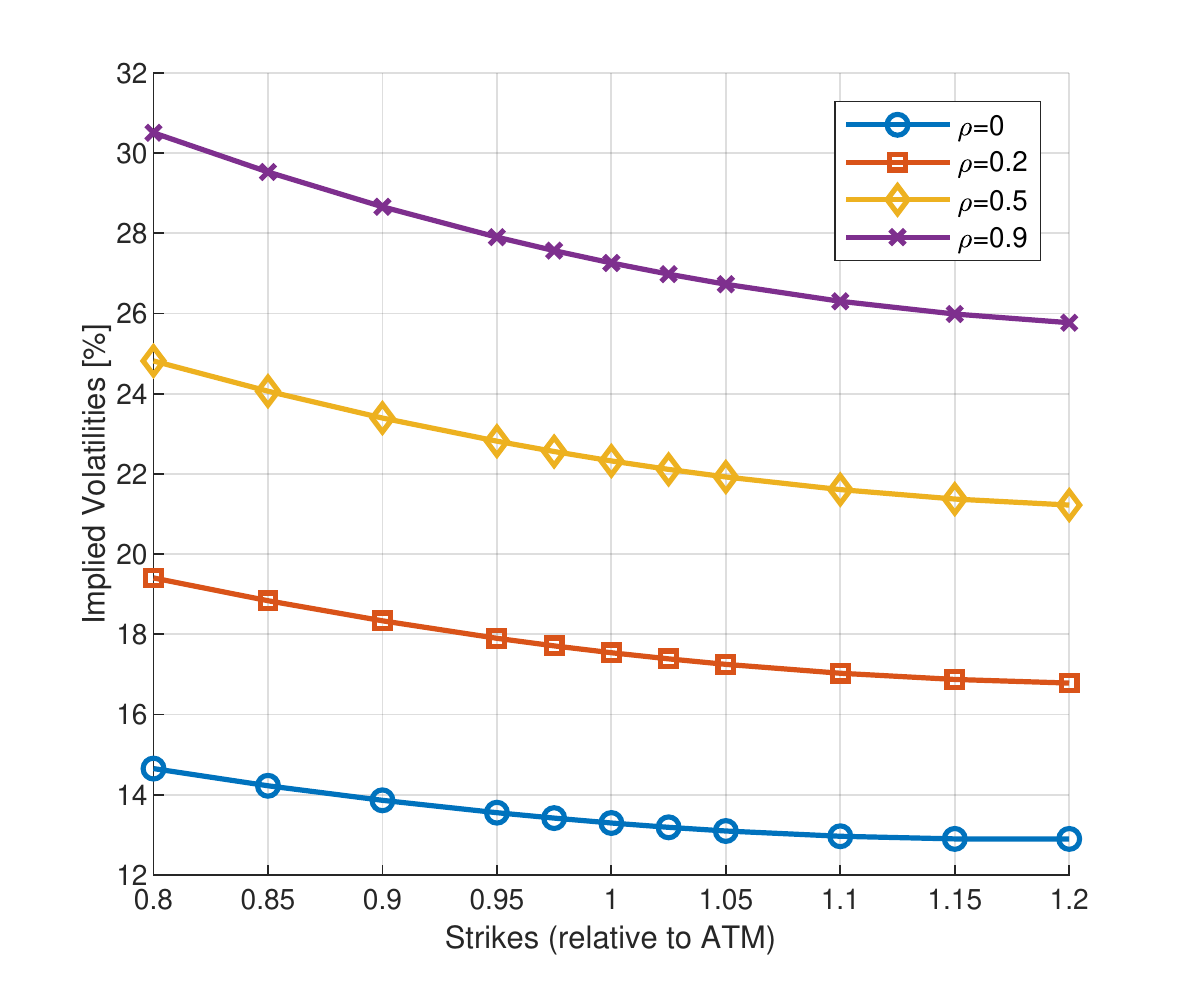}
    \caption{Impact of different jump parameters on the shape of the basket implied volatility in the ELV model. First figure: impact of $\xi_p$; Second figure: impact of $\rho$- correlation between Brownian motions.}
      \label{fig:impact2}	
\end{figure}

\begin{figure}[h!]
  \centering
    \includegraphics[width=0.55\textwidth]{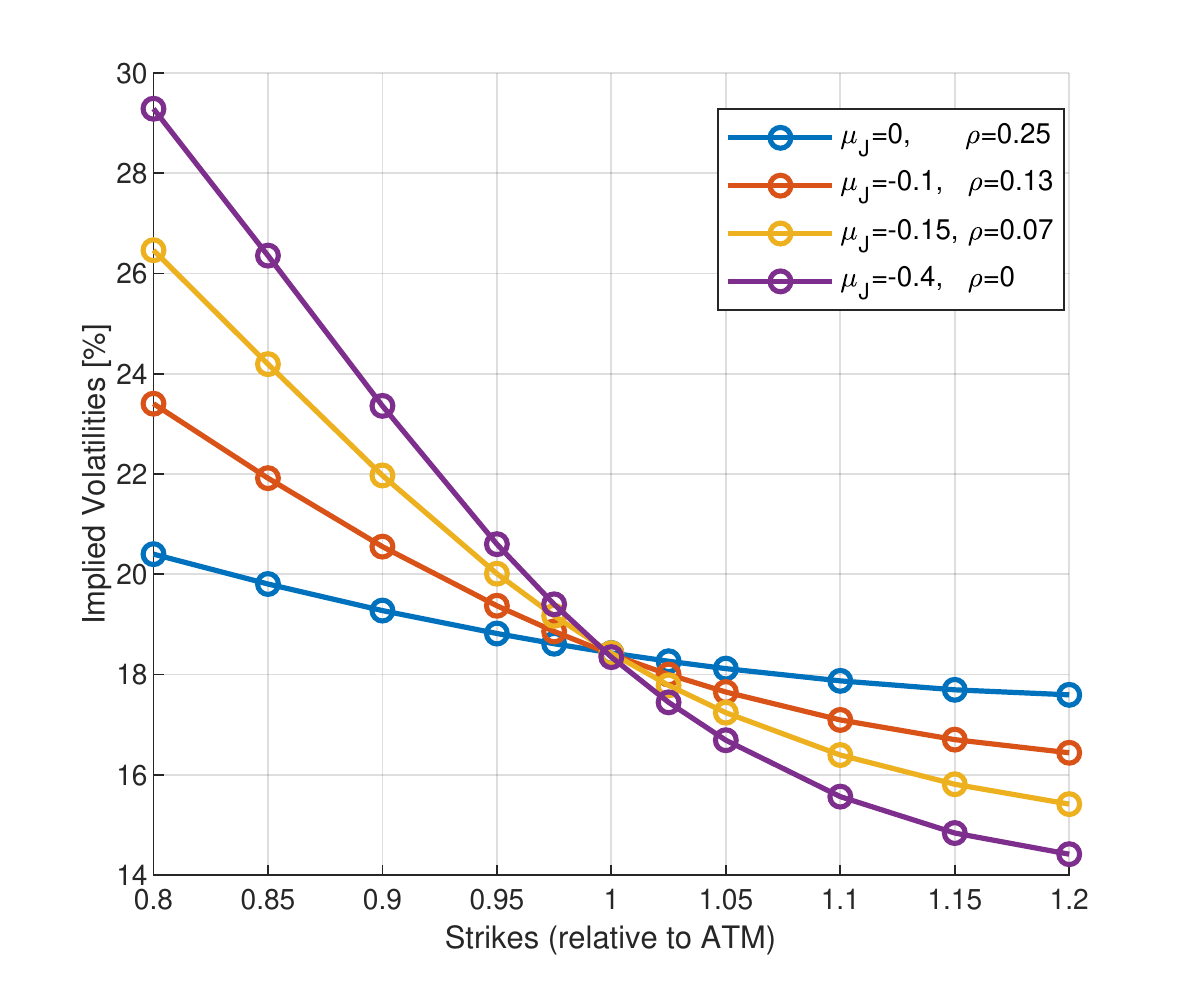}
    \caption{Basket implied volatilities with re-calibrated model parameters for ATM level.}
      \label{fig:impactCombined}	
\end{figure}

Figure~\ref{fig:impact1} and Figure~\ref{fig:impact2} show that\footnote{Results for $\sigma$ are not presented here as they resembled the impacts of $\xi_p$ and $\sigma_J$}: $\sigma$, $\sigma_J$ and $\xi_p$ have a moderate impact on the level of basket implied volatilities. The two remaining parameters are the most relevant: $\mu_J$ and $\rho$. To a large extent, correlation, $\rho$ determines the level of basket implied volatilities (higher correlation lower the implied volatility) while $\mu_J$ in a significant way controls the implied volatility skew, as demonstrated in Figure~\ref{fig:impactCombined}. We considered negative values for $\mu_J$ in the experiment, implying negative shocks to the underlying process. We observe that a more negative value of $\mu_J$ produces more skew.

Since both model parameters $\sigma_J$ and $\xi_p$ impact the basket implied volatility in a similar fashion, this means that in the calibration routine, one of them can be effectively fixed, reducing the complexity of the calibration procedure. First, the calibration process can be performed in an iterative way, where the $\mu_J$ (skew) is chosen; secondly, the ATM implied volatility is found such that the basked and index ATM volatility matched perfectly.

The presented results suggest that the model can only generate implied volatility skew. The basked implied volatility shapes may also be affected by the basket's composition. In Figure~\ref{fig:JumpImpactSingleAsset} we have shown that the kernel process allows for the generation of a smile, though limited. On the other hand, each asset's volatility contributes to the overall basket vol. This will be confirmed in the numerical experiment where we will consider a DJIA index with 30 underlying assets.

\subsection{Calibration of a High Dimensional Basket}
\label{sec:CalibrationBasket}
In this section, we perform a practical experiment where we build a bridge between a basket and an index. This section considers a DJIA (Dow Jones Industrial Average) with 30 underlying assets, as of 12/08/2021, with the spot price of $357.55$. This experiment poses an extension of a case considered in Section~\ref{sec:5Dexample} where a 5D case was examined. Now, however, we will also perform the model calibration, allowing us for consistent pricing of both an index and a combination of basket constitutes.

The essence of the ELV model lies in the construction of a local volatility model that is constructed of basket marginal distributions that are consistent with the index. As such, the calibration procedure can be performed separately for each expiry date $T_i$, where we utilize analytic expression for the basket's moments (see Equation~(\ref{eqn:opitmization2}) that are matched with the moments computed from the index. The calibration is straightforward, and as indicated earlier, there are two dominant model parameters, $\mu_J$ and $\rho$, that play a central role in the skew calibration. Given the closed-form formulas for the moments, the calibration is swift and straightforward and can be performed with a simple search algorithm.

In Table~\ref{tbl:calib_results} the estimated model parameters are presented, while the moments of the basket compared to the index and ``exact'' moments obtained from Monte Carlo are tabulated in Table~\ref{tbl:MomentsBasketIndex}.

\begin{table}[!h]
\caption{Calibrated model parameters.}
\label{tbl:calib_results}
\centering\footnotesize
\begin{tabular}{c|c|c|c|c|c|c}
  \hline
 $T_i$& $\sigma_M$&$\sigma_J$&$\mu_J$&$\xi_P$&$T$&$\rho$\\\hline\hline
 $1m$&0.25&0.020&-0.20&0.25&0.30&0.35\\
 $2m$ &0.15&0.020&-0.20&0.25&0.50&0.25\\
 $3m$  &0.15&0.020&-0.20&0.25&0.50&0.25\\
 $6m$   &0.15&0.020&-0.20&0.25&0.50&0.37\\
 $1y$     &0.1&0.020&-0.20&0.25&0.50&0.45\\
 $1.5y$&0.1&0.020&-0.20&0.25&0.50&0.45\\
 $2y$& 0.20&0.020&-0.25&0.50&0.45&0.38\\ \hline
\end{tabular}
\end{table}

\begin{table}[!h]
\caption{Moment approximation quality for DJIA with 30 underlying assets, as defined in Table~\ref{tbl:Moments}. } \label{tbl:MomentsBasketIndex}
\centering\footnotesize
\begin{tabular}{c|c|c||c|c||c|c}
&\multicolumn{2}{c||}{Index (Market)}&\multicolumn{2}{c||}{Basket (Monte Carlo)}&\multicolumn{2}{c}{Basket (Proxy)}\\ \hline
$T_i$& $\sigma_{B}$& $\E[B^3]/\sigma_B^3$& $\sigma_{B}$&$\E[B^3]/\sigma_B^3$&$\sigma_{B}$&$\E[B^3]/\sigma_B^3$\\\hline\hline
$1m$& 19.08&	6,634.4&19.0& 	 6,714.1& 	 18.7& 	 6,982.6 \\
$2m$& 29.07	&1,895.9&29.3& 	 1,858.0& 	 28.6& 	 1,958.6 \\
$3m$& 36.47	&969.8&36.0& 	 1,009.5& 	 35.2& 	 1,059.5 \\
$6m$& 53.34	&319.9&54.3& 	 304.0& 	 53.7& 	 302.8 \\
$1y$& 78.56	&107.3&81.9& 	 95.9& 	 82.8& 	 85.8 \\
$1.5y$& 94.52	&65.3&99.0& 	 58.0& 	 99.7& 	 50.6 \\
$2y$& 109.07	&45.4&113.5& 	 41.3& 	 113.1& 	 35.5 \\
\end{tabular}
\end{table}

\begin{figure}[h!]
  \centering
    \includegraphics[width=0.48\textwidth]{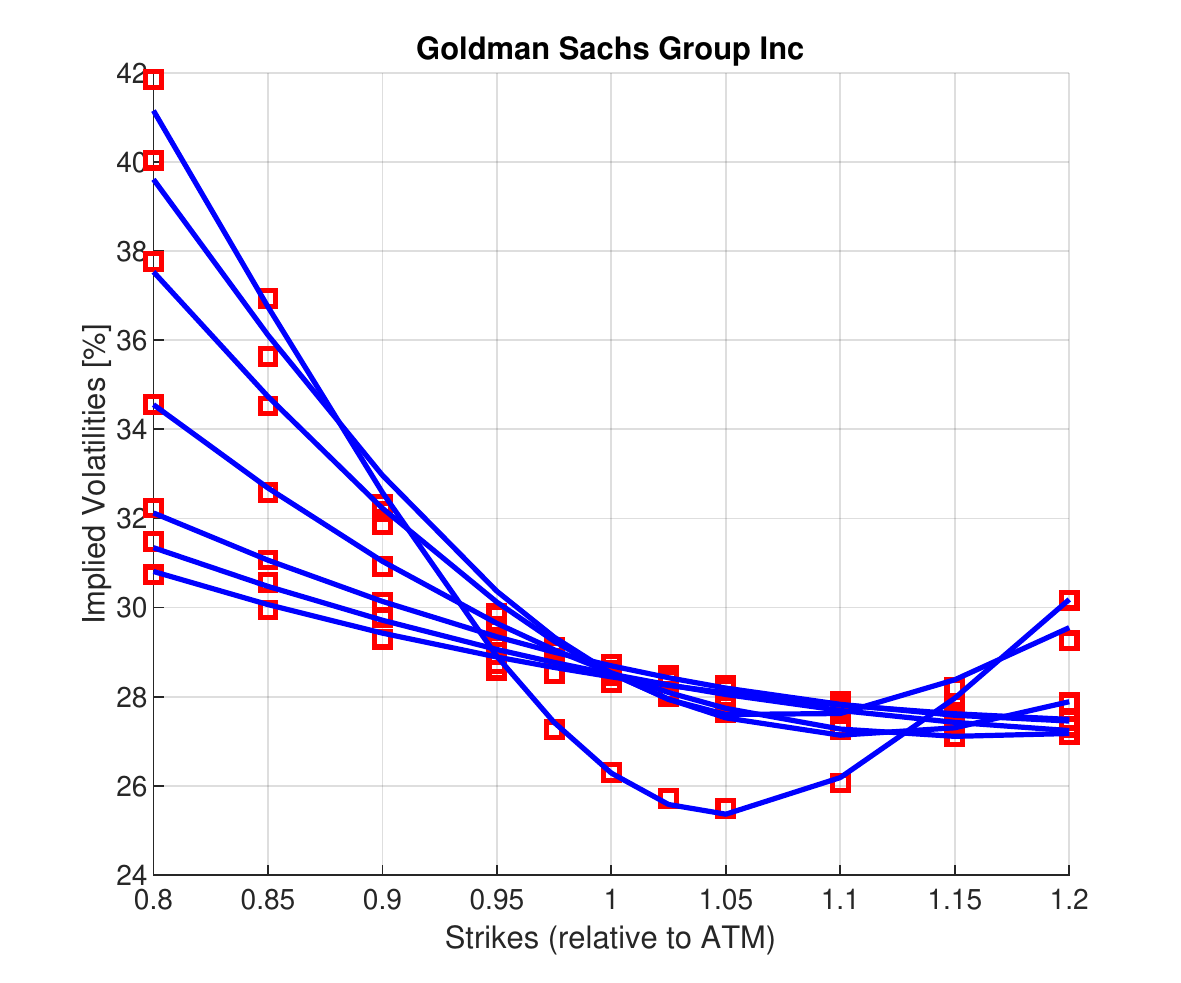}
    \includegraphics[width=0.48\textwidth]{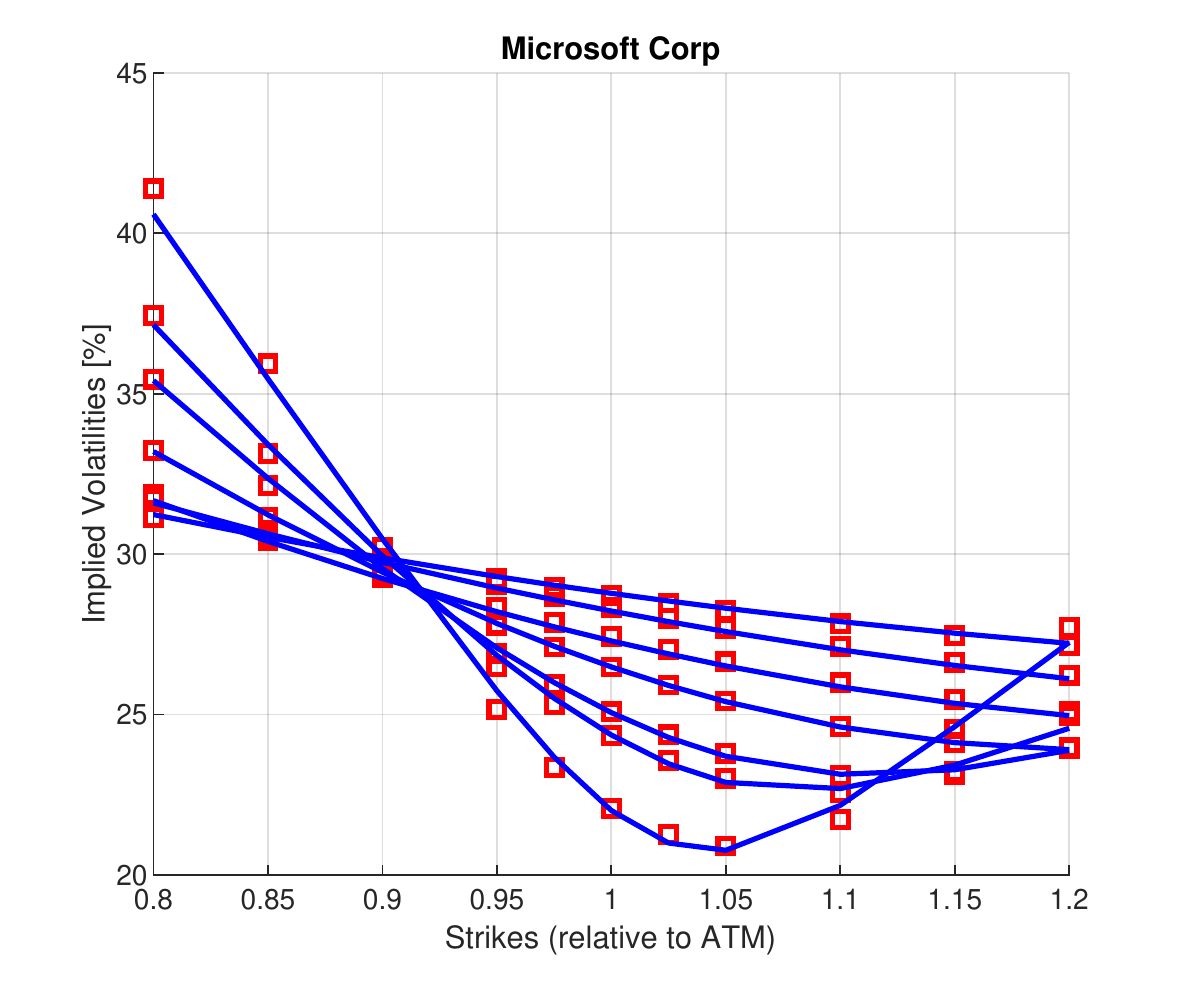}\\
    \caption{Calibration quality for four selected stocks. Red squared indicate market quote, blue line corresponds to the collocation method.}
    \label{fig:marginals}
\end{figure}
\begin{figure}[htb!]
\centering
 \includegraphics[width=0.48\textwidth]{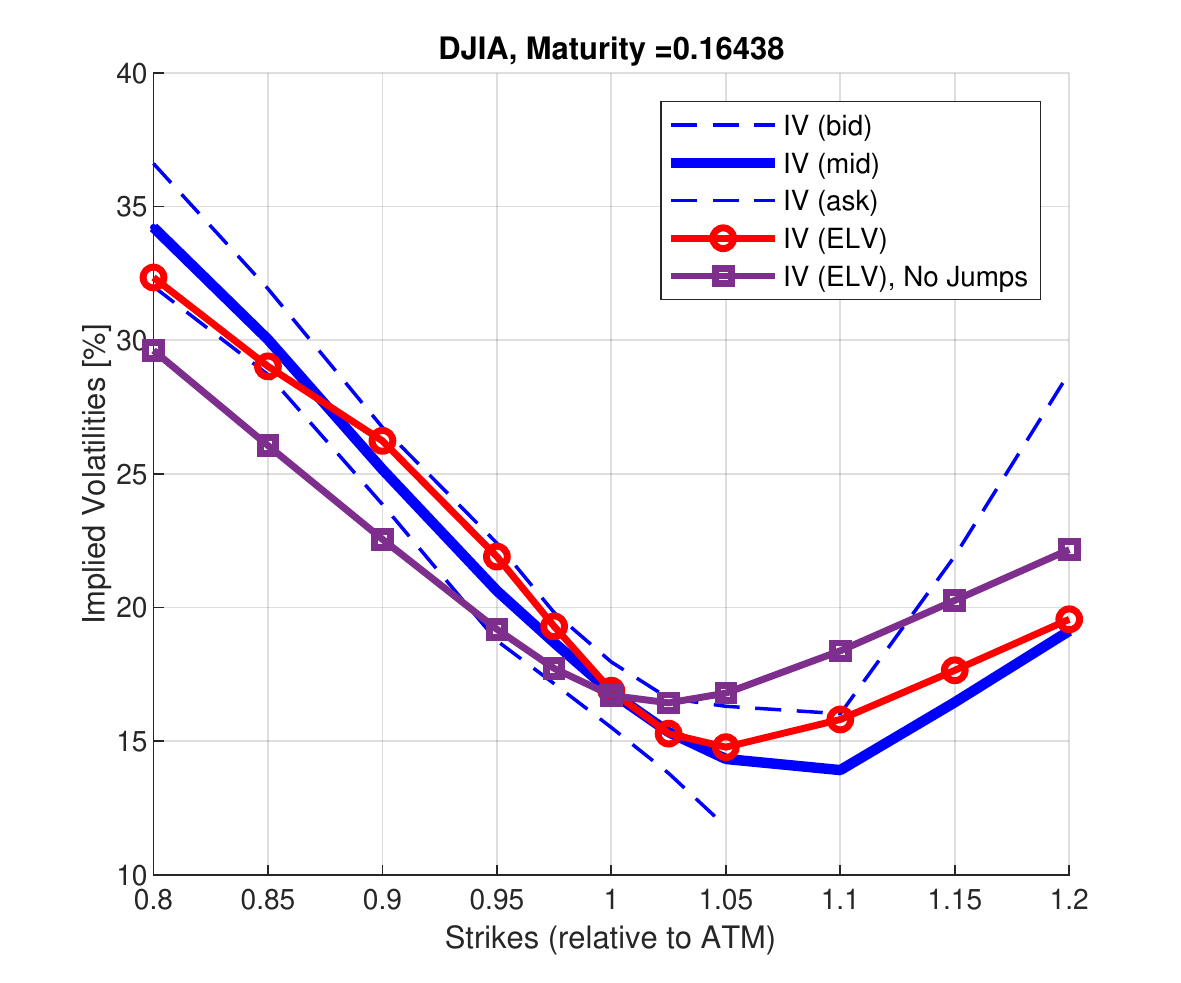}
 \includegraphics[width=0.48\textwidth]{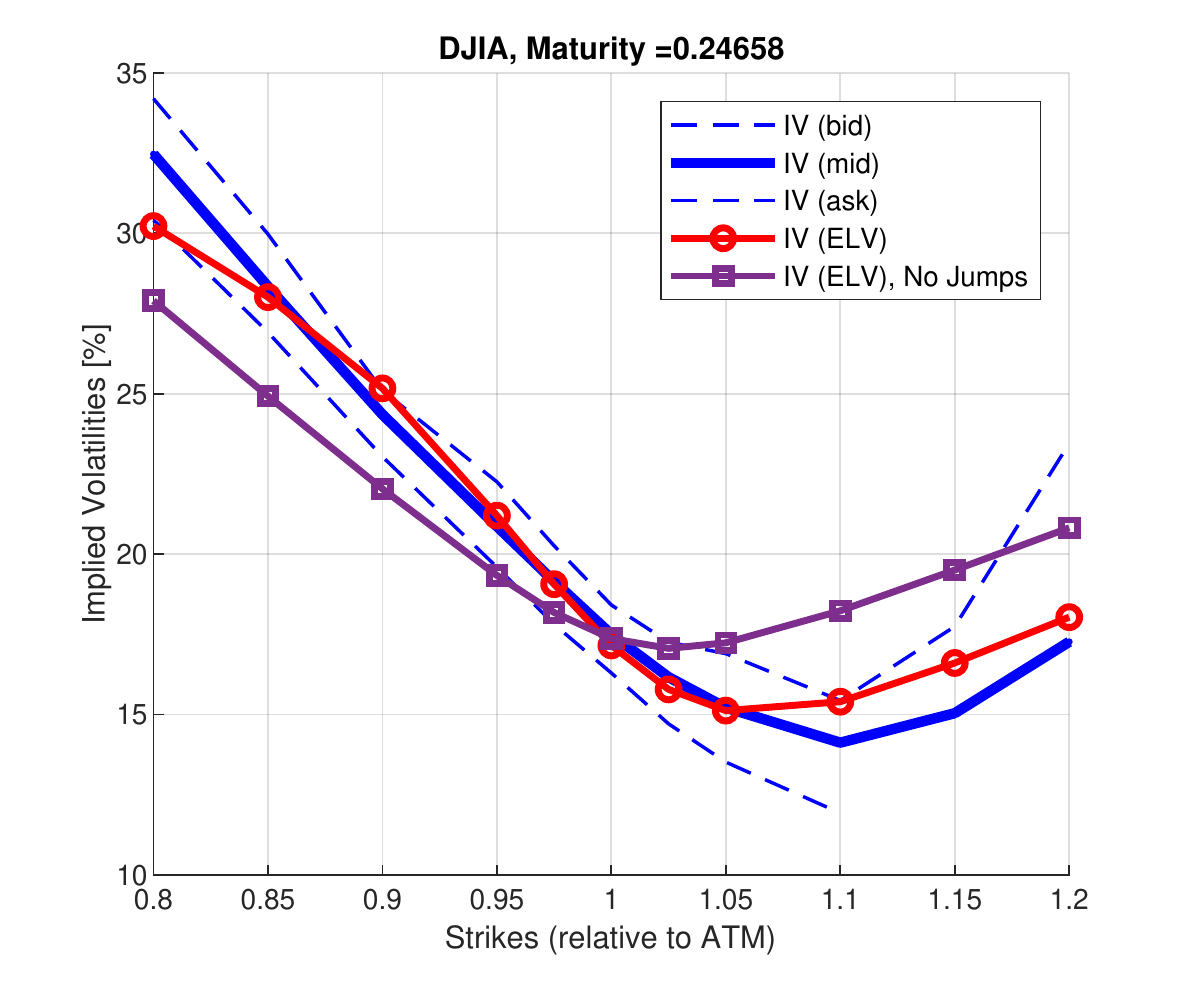}
\caption{Implied volatilities for DJIA 30 compared to ELV model with and without jumps; LHS: $T=2M$, RHS: $T=3M.$}
\label{fig:BasketCalibration1}
\end{figure}
\begin{figure}[htb!]
\centering
 \includegraphics[width=0.48\textwidth]{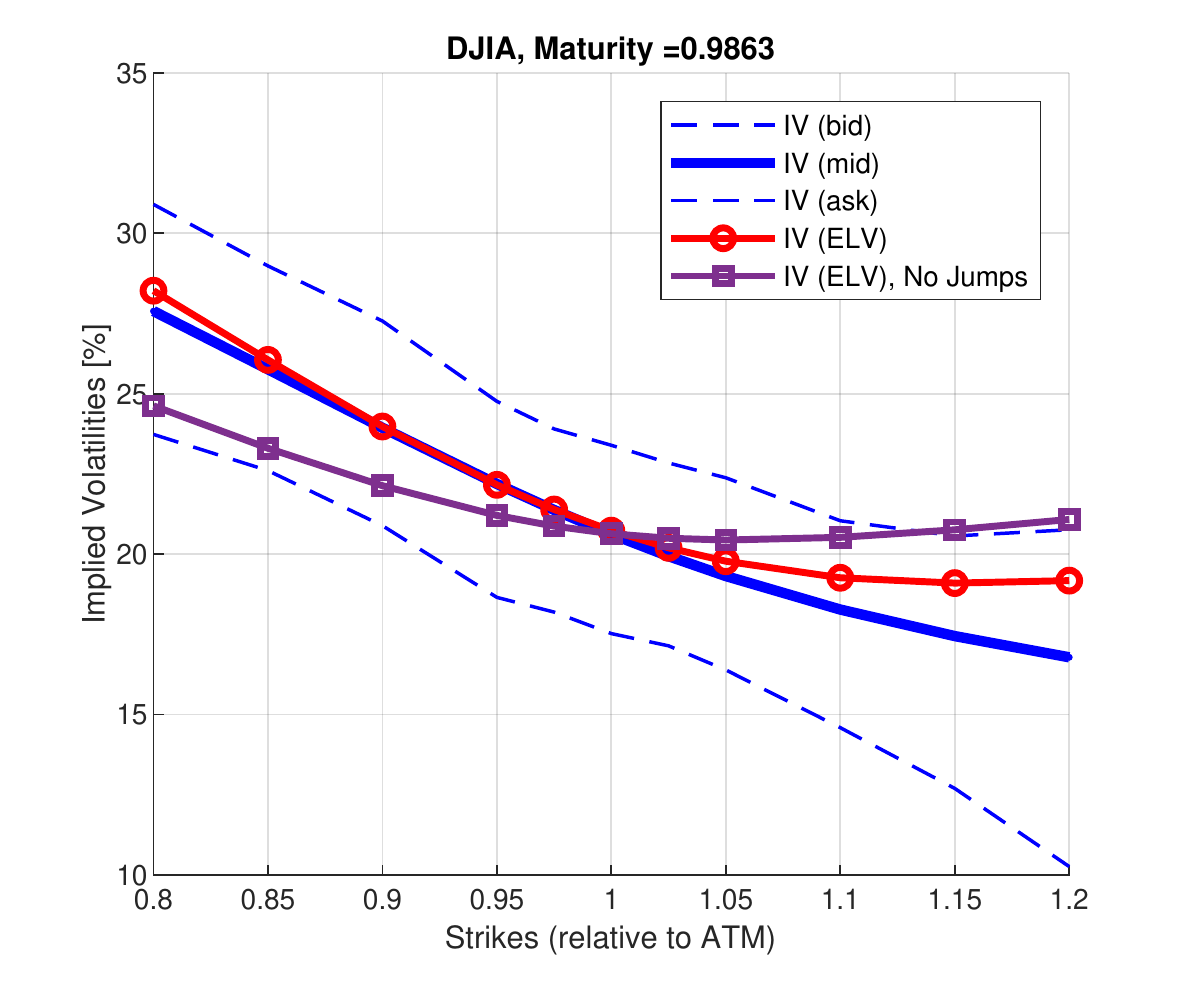}
 \includegraphics[width=0.48\textwidth]{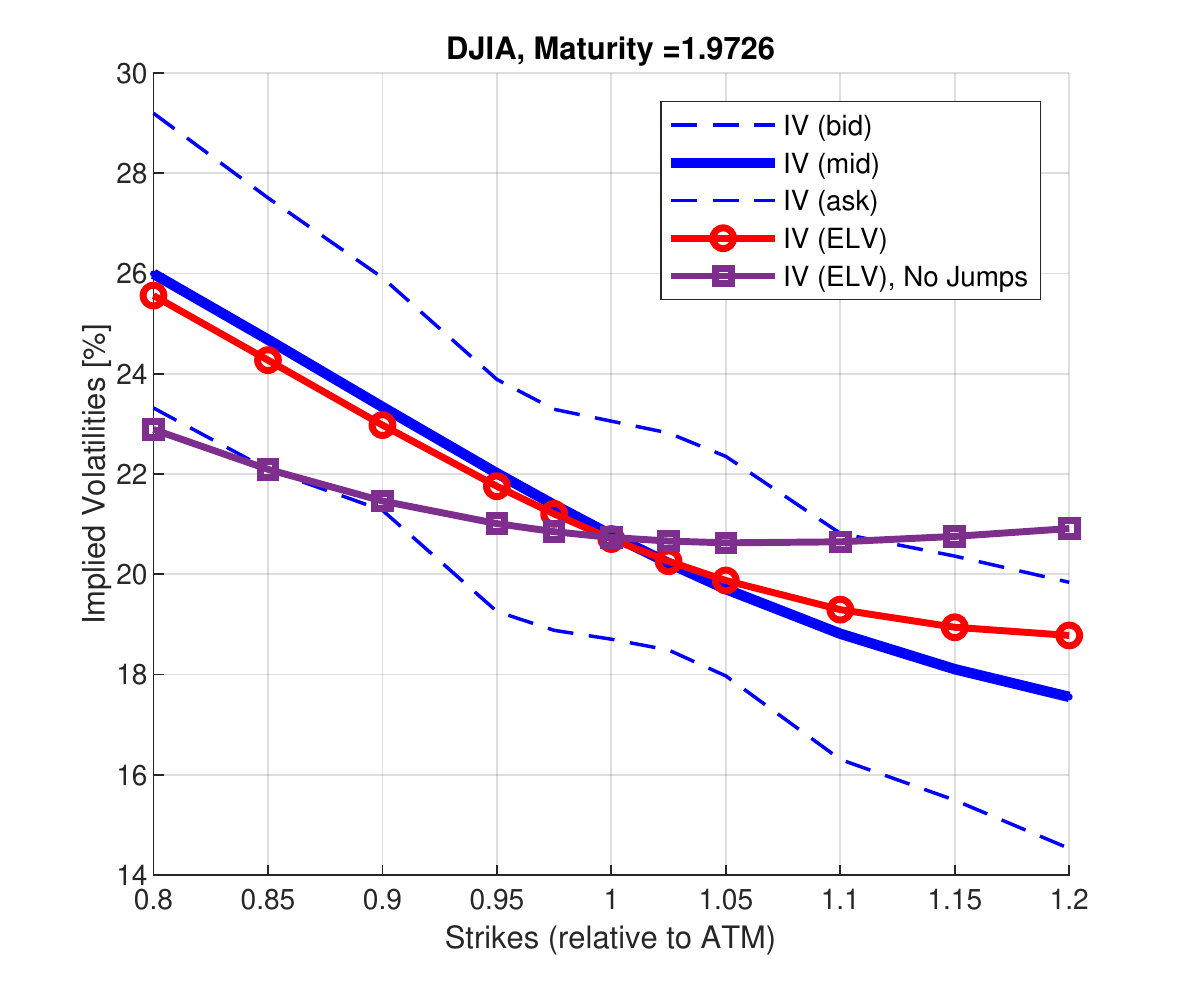}
\caption{Implied volatilities for DJIA 30 compared to ELV model with and without Jumps; LHS: $T=1y$, RHS: $T=2y.$}
\label{fig:BasketCalibration2}
\end{figure}

Figures~\ref{fig:BasketCalibration1} and~\ref{fig:BasketCalibration2} show, for a range of maturities, the implied volatilities obtained from the calibrated basket and the index. In addition, Figure~\ref{fig:marginals} zooms in on the fit to marginal distributions of a selected four stocks included in the index.

The results confirm excellent performance of the ELV model in fitting both individual stocks' distributions and their covariance structure. For all the available index quotes, the model is within the bid-ask spreads. We have also included the ELV model, where jumps are not included for comparison reasons. Such a case may be considered standard local volatility. The additional degree of freedom given by the jumps is essential in calibrating the basket to the index. This is also confirmed by the calibration results shown in the Table~\ref{tbl:calib_results} where negative $\mu_J$ was obtained for all the option expiries.

\subsection{Cost of Basket Miscalibration}
In this section, we assess the impact of basket miscalibration. In the experiment, we will continue with the 30-dimensional basket calibrated in Section~\ref{sec:CalibrationBasket}. We will consider a subset of assets $S_{i_1},\dots S_{i_K}$ of the basket and price an exotic derivative using two variants of the ELV framework. In the first case, we take the model with a covariance structure only driven by Brownian motion, while the alternative is the model where jumps are included. Our previous experiment has shown that including jumps facilitates a better fit to the index. Now, we quantify the potential cost of this mispricing.

In the first step of pricing, the local volatility surface for the ELV dynamics, $\bar{B}$ in~(\ref{eqn:LV_S}), needs to be constructed. This is performed based on the calibrated basket in Equation~(\ref{eqn:BasketProxy}). The number of expiries, $N_T$, available for building the local volatility surface depends on the required accuracy and discretization time-step of~(\ref{eqn:LV_S}). Although, in practice, a finite and somewhat limited set of expiries, $T_j$, is available in the market, for the intermediate expiry, $T$, $T_j<T<T_{j+1}$ we can impose a smooth transition between the corresponding model parameters, i.e.,
\[{\bf p}(T)=\frac{T_{j+1}-T}{T_{j+1}-T_j}{\bf p}(T_j)+\frac{T-T_{j}}{T_{j+1}-T_j}{\bf p}(T_{j+1}),\]
where ${\bf p}(T_j)$, ${\bf p}(T_{j+1})$ represent the parameters of the kernel process $X(T_j)$, $X(T_{j+1})$ observed at times $T_j$ and $T_{j+1}$, respectively.

In this experiment we consider a basket consisting of $5$ stocks, already introduced earlier in Section~\ref{sec:5Dexample} where the subset of DJIA stocks has been considered, and two exotic derivatives, namely, an arithmetic Asian option,
 defined as:
 \[V_{A}(t_0)=N\e^{-rT}\E\left[\left(\sum_{l=1}^{L}S(T_l)-K_A\right)^+|\F(t_0)\right],\]
 with $N$ being the notional and where $K_A$ is the strike price, and the up-and-out barrier option:
\[V_{Bar}(t_0)=N\e^{-rT}\E\left[(B(T)-K_B)^+1_{\tau(H,B)>T}\Big|\F(t_0)\right],\]
where $K_B$ is the strike price, $H>B(t_0)$ is the barrier, and $\tau(x,B):=\inf\{t\geq0:B(t)>x\}$, and the notional, $N$. In the experiment we consider the ATM option with $K_B=123.82$ and the notional of $1$.
Although relatively standard in the financial world, both derivatives are sensitive to different characteristics of the underlying model.

\begin{figure}[h!]
  \centering
    \includegraphics[width=0.48\textwidth]{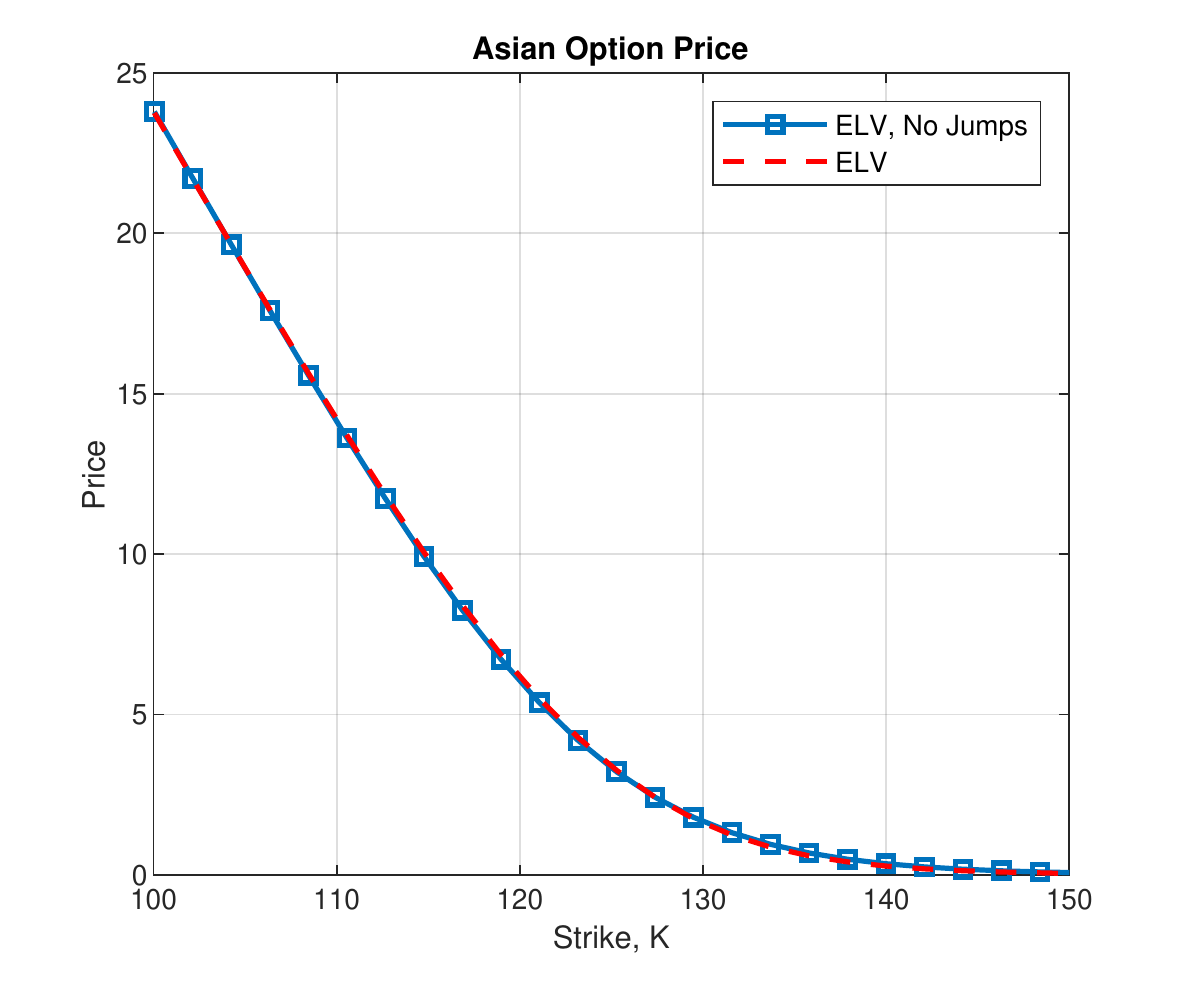}
    \includegraphics[width=0.48\textwidth]{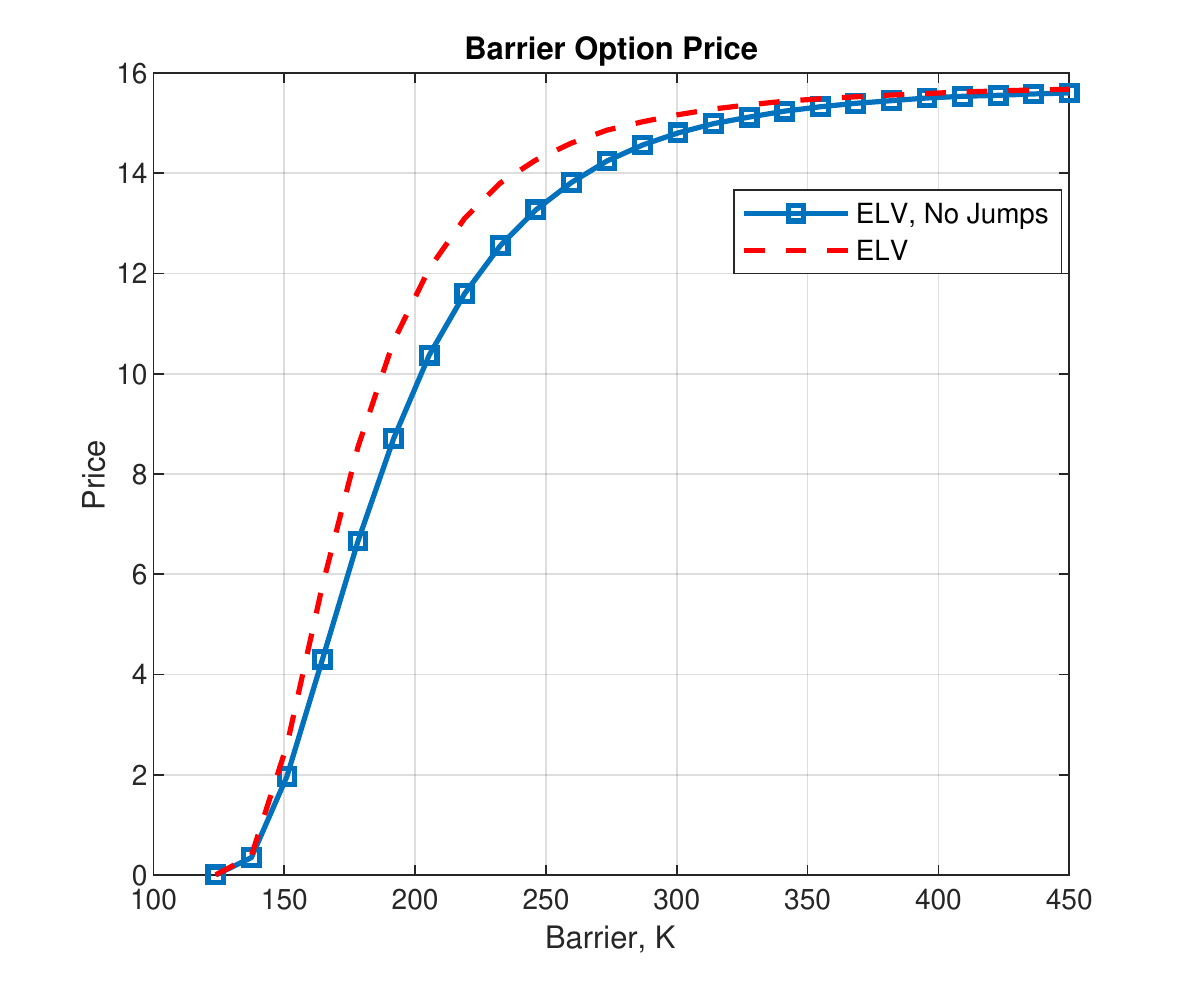}\\
    \caption{Figure illustrates prices of Asian and Barrier option prices for: LHS varying strike, $K$, RHS: varying the barrier level, $H$.}
    \label{fig:AsianAndBarrier}
\end{figure}

The pricing for varying strikes and barrier levels is presented in Figure~\ref{fig:AsianAndBarrier}. The results show that although in the case of the Asian option, the skew-related miscalibration does not have a significant impact in the case of the up-and-out barrier option, the skew plays a vital role. As illustrated in Table~\ref{tbl:AsianBarrier} the price differences may be even ten-fold.

\begin{table}[!h]
\caption{Comparison of Asian and Barrier option pricing for ELV models with and without jumps.}
\label{tbl:AsianBarrier}
\centering\footnotesize
\begin{tabular}{c|c|c|c|c|c|c|c}
  \hline
  \multicolumn{3}{c}{Asian option price differences}\\\hline
 $K$&100.05&	108.46&	116.87&	125.28&	133.69&	142.10&	150.51\\
 diff&0.0392&	-0.0555&	-0.1439&	-0.0464&	0.0870&	0.0602&	0.0289\\ \hline
 \multicolumn{3}{c}{Barrier option price differences}\\\hline
 H&123.8225&	178.1855&	232.5484&	286.9113&	341.2742&	395.6371&	450\\
 diff&0&	-1.881&	-1.2491&	-0.4639&	-0.1986&	-0.0891&	-0.0735\\
\end{tabular}
\end{table}

\section{Conclusions}\label{conclusions}
We have proposed above the Effective Local Volatility model capable of accurate, consistent and swift pricing of basket options -- both plain vanilla and of the more exotic type. The model is accurate, because -- as demonstrated in Section~\ref{sec:CalibrationBasket} -- it calibrates to liquid index options with only minimal error, well within the bid-ask spread. The model is consistent, because it simultaneously calibrates to individual basket constituents (via the functions $g_{i,j}(\cdot)$) and their index-implied covariance structure (via the kernel processes $X_i(\cdot)$ which we assume to be of Merton's jump diffusion type). Finally, the model works swiftly, because the typically numerically ``expensive'' task, like sampling from stocks' marginal distributions, is handled via the collocation method, and basket calibration is performed by means of moment matching technique, leveraging the analytical expressions for basket moments which we derive in Section~\ref{sect_calibration}. Importantly, unlike copula-based or local-correlation models which tend to be black-boxy and intractable, the proposed approach allows users to control the basket implied volatility skew using just a few parameters in a clear and transparent way. Thus, it is straightforward to control the dependence structure of the basket and trace how implied volatility smiles transfer from individual assets to the whole basket (or possibly the index). While interesting in its own right, this feature makes it possible to price structured products including baskets on any combination of index constituents stocks: the assets simply ``inherit'' the covariance structure implied by the index which we then feed into a one-dimensional local volatility model, allowing us to price a desired structured product in a quick and easy way.

While the model as currently proposed strikes a good balance between accuracy and tractability and works very well for basket dimensions relevant in practice, the approach can nonetheless be extended in two main directions. First, one could postulate more sophisticated dynamics for the kernel processes than the simple jump diffusion model. This might allow for even better control of the implied volatility shapes, however possibly at some cost to tractability, slowing down the calibration. Second, one could add a stochastic component to the Effective Local Volatility model which might allow for more accurate pricing of callable structures.

\bibliography{article}
\bibliographystyle{abbrv}


\appendix

\section{Stochastic Collocation Method}\label{sc:scmc}
Let's assume two random variables, $Y$ and $X$, where the latter one is cheaper to sample from (e.g.,  $X$ is a Gaussian random variable).
These two scalar random variables are connected, via,
\begin{equation}
    F_Y(Y) \stackrel{d}{=} U \stackrel{d}{=} F_X(X),
\end{equation}
where  $U\sim \mathcal{U}([0,1])$ is a uniformly distributed random variable, $F_Y(\bar y):=\P[Y \leq \bar y]$ and $F_X(\bar x):=\P[X \leq \bar x]$ are cumulative distribution functions (CDF).  Note that $F_X(X)$ and $F_Y(Y)$ are random variables following the same uniform distribution. $F_Y(\bar y_n)$ and $F_X(\bar x_n)$ are supposed to be strictly increasing functions, so that the following inversion holds true,
\begin{equation} \label{eq:xydist}
    \bar y_n = F^{-1}_Y(F_X(\bar x_n))=:g(\bar x_n)  .
\end{equation}
where $\bar y_n$ and $\bar x_n$ are samples (numbers) from $Y$ and $X$, respectively. The mapping function, $g(\cdot) =  F^{-1}_Y(F_X(\cdot))$, connects the two random variables and guarantees that $F_X(\bar x_n)$ equals $F_Y(g(\bar x_n))$, in distributional sense and also element-wise. The mapping function should be approximated,  i.e., $g(\bar x_n)\approx g_m(\bar x_n)$, by a function which is cheap. When function $g_m(\cdot)$ is available, we may generate ``expensive'' samples, $\bar y_n$ from $Y$, by using the cheaper random samples $\bar x_n$ from $X$.

The Stochastic Collocation Monte Carlo method (SCMC) developed in~\cite{Grzelak:2015SCMC} aims to find an accurate mapping function $g(\cdot)$ in an efficient way.
The basic idea is to employ Equation~\eqref{eq:xydist} at specific collocation points and approximate the function $g(\cdot)$ by a suitable monotonic interpolation between these points.
This procedure, see Algorithm~I, reduces the number of expensive inversions $F^{-1}_Y(\cdot)$ to obtain many samples from $Y(\cdot)$.

\zbox{
{\bf Algorithm: SCMC Method} \\
\label{alg2}
Taking an interpolation function of degree $m-1$ (with $m\geq2$, as we need at least two collocation points), as an example, the following steps need to be performed:
\begin{itemize}
\setlength\itemsep{0.0em}
	\item[1.]  Calculate CDF $F_X(x_j)$ on the points $(x_1,x_2, ..., x_m)$, that are obtained, for example, from Gauss-Hermite quadrature, giving $m$ pairs $({x}_j, F_X({x}_j))$;
	\item[2.]  Invert the target CDF   ${y}_j=F^{-1}_Y(F_X({x}_j))$, $j=1,\ldots, m$, and form $m$ pairs $({x}_j, {y}_j)$;
	\item[3.]  Define the interpolation function, $y=g_m(x)$, based on these $m$  point pairs $({x}_j,{y}_j)$;
    \item[4.] Obtain sample $\hat{Y}$ by applying the mapping function $\hat{Y}=g_m(\hat{X})$, where  sample $\hat{X}$ is drawn from $X$.
\end{itemize}
}

The SCMC method parameterizes the distribution function by imposing probability constraints at the given collocation points.  Taking the Lagrange interpolation as an example, we can expand function $g_m(\cdot)$ in the form of polynomial chaos,
\begin{equation} \label{eq:scmc-lang}
    Y \approx g_m(X) = \sum_{j=0}^{m-1}\hat\alpha_{j} X^{j} = \hat\alpha_0+\hat\alpha_1X  + ... + \hat\alpha_{m-1} X^{m-1}.
\end{equation}
Monotonicity of interpolation is an important requirement, particularly when dealing with peaked probability distributions.

The Cameron-Martin Theorem~\citep{Cameron1947AnnMath}  states that any distribution can be approximated by a polynomial chaos approximation based on the normal distribution, but also other random variables may be used for $X$ (see, for example,~\cite{Grzelak:2015SCMC}).

\section{Proof of Lemma~\ref{lem:Merton2d}}
\label{app:proofMerton2D}
\begin{proof} By substitution we find:
\begin{eqnarray*}
\E[\e^{aX_{1}(T)+bX_{2}(T)}]=\e^{a\mu_1+b\mu_2}\E\left[\exp\left( a\sigma_1W_{1}(T)+  b \sigma_2W_{2}(T)+\sum_{k=1}^{X_{\mathcal{P}}(T)} aJ_{k,{1}}+\sum_{k=1}^{X_{\mathcal{P}}(T_j)}bJ_{k,2}\right)\right],
\end{eqnarray*}
with $X_{1}$, $X_{2}$ defined in~(\ref{eqn:2DMarton}), and where $J_{k,i}\sim\mathcal{N}(\mu_J,\sigma^2_J)$.
Due to the correlation between Brownian motions we perform the following factorization,
\[a\sigma_1W_{1}(T)+  b\sigma_2 W_{2}(T)\stackrel{\text{d}}{=}\sqrt{a^2\sigma_1^2+b^2\sigma_2^2+2\rho_{1,2}ab\sigma_1\sigma_2}\widetilde{W}(T)=:\hat\sigma\widetilde{W}(T),\]
with $\hat\sigma^2=a^2\sigma_1^2+b^2\sigma_2^2+2\rho_{1,2}ab\sigma_1\sigma_2$ and where $\widetilde{W}(T)$ is an independent Brownian motion.
\begin{eqnarray*}
\sum_{k=1}^{X_{\mathcal{P}}(T)} aJ_{k,{1}}+\sum_{k=1}^{X_{\mathcal{P}}(T)}bJ_{k,2}=\sum_{k=1}^{X_{\mathcal{P}}(T)}\left(aJ_{k,{1}}+bJ_{k,{2}}\right)=:\sum_{k=1}^{X_{\mathcal{P}}(T)} \hat J_{k},
\end{eqnarray*}
where
$\hat J\equiv \hat J_{k}=aJ_{k,{1}}+bJ_{k,{2}}\sim \mathcal{N}((a+b)\mu_J,(a^2+b^2)\sigma_J^2)$; thus we have:
\begin{eqnarray*}
\E[\e^{aX_{1}(T)+bX_{2}(T)}]=\E\left[\exp\left( \hat\sigma\widetilde{W}(T)+\sum_{k=1}^{X_{\mathcal{P}}(T)} \hat J_{k}\right)\right]\stackrel{\perp \!\!\! \perp}{=}\e^{a\mu_1+b\mu_2}\E\left[\e^{\hat\sigma\widetilde{W}(T)}\right]\E\left[\e^{\sum_{k=1}^{X_{\mathcal{P}}(T)} \hat J_{k}}\right].
\end{eqnarray*}
The first expectation simply reads
$\E\big[\e^{\hat\sigma\widetilde{W}(T)}\big]=\e^{\frac12\hat\sigma^2T},$ while the second expectation, thanks to the tower property of expectations, gives:
\begin{eqnarray*}
\E\left[\exp\left(\sum_{k=1}^{X_{\mathcal{P}}(T)} \hat J_{k}\right)\right]&=&\sum_{n\geq0}\E\left[\exp\left(\sum_{k=1}^n \hat J_{k}\right)\right]\frac{\e^{-\xi_pT(\xi_pT)^n}}{n!}\\
&=&\sum_{n\geq0}\frac{\e^{-\xi_pT(\xi_pT)^n}}{n!}\E^n\left[\e^{\hat J}\right]\\
&=&\exp\left(\xi_pT\E[\e^{\hat J}-1]\right),
\end{eqnarray*}
where the last expression can be recognized as a Taylor expansion of an exponential function. Since \[\E[\e^{\hat J}]=\e^{(a+b)\mu_J+\frac12(a^2+b^2)\sigma_J^2},\]
by substitutions the proof is complete.
\end{proof}

\section{Third Moment Derivations}
\label{appendix:thirdmoment}
For convenience we neglect the time argument $T_j$. The third moment of the basket, by definition, is defined as:
\begin{eqnarray*}
\E[B^3]=\sum_{i_1=1}^N\sum_{i_2=1}^N\sum_{i_3=1}^N\omega_{i_1}\omega_{i_2}\omega_{i_3}\E\left[S_{i_1}S_{i_2}S_{i_3}\right]=:\sum_{i_1=1}^N\sum_{i_2=1}^N\sum_{i_3=1}^N\omega_{i_1}\omega_{i_2}\omega_{i_3}\E\left[S_{i_1,i_2}S_{i_3}\right],
\end{eqnarray*}
where $S_{i_1,i_2}=S_{i_1}S_{i_2}$.
Following the same procedure as for the second moment we find:
\begin{eqnarray*}
\E[B^3]&=&\sum_{i_1=1}^N\sum_{i_2=1}^N\sum_{i_3=1}^N\omega_{i_1}\omega_{i_2}\omega_{i_3}\left[\hat\rho({S_{i_1,i_2},S_{i_3}})\sigma_{S_{i_1,i_2}}\sigma_{S_{i_2}}+\E[S_{i_1,i_2}]\E[S_{i_3}]\right].
\end{eqnarray*}
$\E[S_{i_1,i_2}]:=\E[S_{i_1}S_{i_2}]$, can be computed with~(\ref{eqn:ES1S2}), $\E[S_{i_3}]$ is given from the market data. Terms we need to establish are: $\sigma_{S_{i_1,i_2}}$ and $\hat\rho({S_{i_1,i_2},S_{i_3}})$. For the first term we have:
\begin{eqnarray*}
\sigma_{S_{i_1,i_2}}^2&=&\E[S_{i_1}S_{i_2}]-\E[S_{i_1}]\E[S_{i_2}],
\end{eqnarray*}
which again can be computed using Equation~(\ref{eqn:ES1S2}).
For the correlation coefficient we have:
\begin{eqnarray*}
\hat\rho({S_{i_1,i_2},S_{i_3}})&=&\rho\left(\e^{\hat\alpha_{i_1,j,0}+\hat\alpha_{i_1,j,1}X_{i_1}}\e^{\hat\alpha_{i_2,j,0}+\hat\alpha_{i_1,j,1}X_{i_2}},\e^{\hat\alpha_{i_3,j,0}+\hat\alpha_{i_3,j,1}X_{i_3}}\right)\\
&=&\rho\left(\e^{\hat\alpha_{i_1,j,1}X_{i_1}+\hat\alpha_{i_2,j,1}X_{i_2}},\e^{\hat\alpha_{i_3,j,1}X_{i_3}}\right)=:\bar\rho({S_{i_1,i_2},S_{i_3}}),
\end{eqnarray*}
which yields:
\begin{eqnarray*}
\bar\rho({S_{i_1,i_2},S_{i_3}})&\stackrel{\text{def}}{=}&\frac{\E[\e^{\hat\alpha_{i_1,j,1}X_{i_1}+\hat\alpha_{i_2,j,1}X_{i_2}+\hat\alpha_{i_3,j,1}X_{i_3}}]-\E[\e^{\hat\alpha_{i_1,j,1}X_{i_1}+\hat\alpha_{i_2,j,1}X_{i_2}}]\E[\e^{\hat\alpha_{i_3,j,1}X_{i_3}}]}{\sqrt{\Var[\e^{\hat\alpha_{i_1,j,1}X_{i_1}+\hat\alpha_{i_2,j,1}X_{i_2}}]\Var[\e^{\hat\alpha_{i_3,j,1}X_{i_3}}]}}\\
&=&\frac{\E[\e^{\hat\alpha_{i_1,j,1}X_{i_1}+\hat\alpha_{i_2,j,1}X_{i_2}+\hat\alpha_{i_3,j,1}X_{i_3}}]-\E[\e^{\hat\alpha_{i_1,j,1}X_{i_1}+\hat\alpha_{i_2,j,1}X_{i_2}}]\phi_{X_{i_3}}(-i\hat\alpha_{i_3,j,1})}{\sqrt{\Var[\e^{\hat\alpha_{i_1,j,1}X_{i_1}+\hat\alpha_{i_2,j,1}X_{i_2}}]\left(\phi_{X_{i_3}}(-2i\hat\alpha_{i_3,j,1})-\phi_{X_{i_3}}^2(-i\hat\alpha_{i_3,j,1})\right)}}.
\end{eqnarray*}
All the expression except for the first expectation can be computed using~(\ref{eqn:ES1S2}):
\begin{eqnarray}
\label{eqn:W3D}
a\sigma_1W_{1}(T)+  b\sigma_2 W_{2}(T)+ c\sigma_3 W_{3}(T)\stackrel{\text{d}}{=}\hat\sigma\widetilde{W}(T),
\end{eqnarray}
with
\begin{eqnarray}
\label{eqn:sigma3D}
\hat\sigma^2 = {a^2\sigma_1^2+b^2\sigma_2^2+c^2\sigma_3^2+2\rho_{1,2}ab\sigma_1\sigma_2}+2\rho_{1,3}ac\sigma_1\sigma_3 +2\rho_{2,3}ab\sigma_2\sigma_3,
\end{eqnarray}
therefore
\begin{eqnarray}
\label{eqn:Jhat3D}
\sum_{k=1}^{X_{\mathcal{P}}(T)} aJ_{k,{1}}+\sum_{k=1}^{X_{\mathcal{P}}(T)}bJ_{k,2}+\sum_{k=1}^{X_{\mathcal{P}}(T)}bJ_{k,3}=\sum_{k=1}^{X_{\mathcal{P}}(T)}\left(aJ_{k,{1}}+bJ_{k,{2}}+bJ_{k,{3}}\right)=:\sum_{k=1}^{X_{\mathcal{P}}(T)} \hat J_{k},
\end{eqnarray}
where
$\hat J\equiv \hat J_{k}=aJ_{k,{1}}+bJ_{k,{2}}+cJ_{k,{3}}\sim \mathcal{N}((a+b+c)\mu_J,(a^2+b^2+c^2)\sigma_J^2)$; thus we have:

\begin{eqnarray}
\E[\e^{aX_{1}+bX_{2}+cX_{3}}]\stackrel{\perp \!\!\! \perp}{=}\e^{a\mu_1+b\mu_2+c\mu_3}\E\left[\e^{\hat\sigma\widetilde{W}(T)}\right]\E\left[\e^{\sum_{k=1}^{X_{\mathcal{P}}(T)} \hat J_{k}}\right],
\end{eqnarray}
with $\hat\sigma\widetilde{W}(T)$ defined in~(\ref{eqn:W3D}) and $\hat J_{k}$ defined in~(\ref{eqn:Jhat3D}).
Using the result from proof in~\ref{app:proofMerton2D} we find:
\begin{eqnarray}
\label{eqn:omega3D}
\omega_X(a,b,c)&:=&\E[\e^{aX_{1}+bX_{2}+cX_{3}}]\\\nonumber
&=&\exp\left(a\mu_1+b\mu_2+c\mu_3+\frac12\sigma T+\xi_pT\E[\e^{\hat J}-1]\right)\\\nonumber
&=&\exp\left[a\mu_1+b\mu_2+c\mu_3+\frac12\hat\sigma T+\xi_pT\left(\e^{(a+b+c)\mu_J+\frac12(a^2+b^2+c^2)\sigma_J^2}-1\right)\right],\nonumber
\end{eqnarray}
where $\sigma^2$ is defined in~(\ref{eqn:sigma3D}).

Combining these results with Lemma~\ref{lem:Merton2d} we find the following expression for the correlation: $\bar\rho({S_{i_1,i_2},S_{i_3}})$,
\footnotesize
\begin{eqnarray*}
\bar\rho({S_{i_1,i_2},S_{i_3}})=\frac{\omega_X(c_1,c_2,c_3)-\omega_X(c_1,c_2)\phi_{X_{i_3}}(-ic_3)}{\sqrt{\left(\omega_X(2c_1,2c_2)-\omega_X^2(c_1,c_2)\right)\left(\phi_{X_{i_3}}(-2ic_3)-\phi_{X_{i_3}}^2(-ic_3)\right)}},
\end{eqnarray*}
\normalsize
where:  $c_1:=\hat\alpha_{i_1,j,1}$, $c_2:=\hat\alpha_{i_2,j,1}$, $c_3:=\hat\alpha_{i_3,j,1}$ and where $\omega_X(c_1,c_2,c_3)$ is defined in~(\ref{eqn:omega3D}) and $\omega_X(c_1,c_2)$ in~(\ref{eqn:Expectation2D}).
\end{document}